\documentclass[11pt,a4paper]{article}
\pdfoutput=1

\usepackage[T1]{fontenc}
\usepackage{amssymb,amsmath}
\usepackage{lmodern}
\usepackage{verbatim,setspace}

\usepackage[utf8]{inputenc}

\DeclareMathAlphabet{\mathpzc}{OT1}{pzc}{m}{it}
\usepackage{dsfont}
\usepackage{amssymb}
\usepackage{amsmath}

\def\cN{{\mathcal{N}}}

\renewcommand\Im{\hbox{{\rm Im}}\,}
\renewcommand\Re{\hbox{{\rm Re}}\,}

\newcommand{\CR}{\nonumber \\*}

\def\cH{{\mathcal H}}
\def\cV{{\mathcal V}}
\newcommand{\ee}{\mathrm{e}}
\def\del{\partial}
\newcommand{\Iprod}[2]{\langle {#1}, {#2} \rangle}
\newcommand{\im}{\mathrm{i}}
\newcommand{\vl}{\mathsf{vol}}
\newcommand{\Jd}[1]{J_{\scriptscriptstyle ({#1})}}
\newcommand{\Od}[1]{\Omega_{\scriptscriptstyle ({#1})}}
\newcommand{\bOd}[1]{\bar\Omega_{\scriptscriptstyle ({#1})}}
\newcommand{\mg}{\mathsf{m}}
\newcommand{\cg}{\mathsf{c}}
\newcommand{\pg}{\mathsf{p}}
\newcommand{\jg}{\mathsf{J}}
\newcommand{\bg}{\mathsf{b}}
\newcommand{\qg}{\mathsf{q}}

\makeatletter
\newcommand{\thickhline}{%
    \noalign {\ifnum 0=`}\fi \hrule height 1.3pt
    \futurelet \reserved@a \@xhline
}

\usepackage{jheppub}

\title{AdS$_4$ black holes from M-theory}
\author[a]{Stefanos Katmadas}  \author[a]{and Alessandro Tomasiello}
\affiliation[a]{Dipartimento di Fisica, Universit\'a di Milano-Bicocca,
I-20126 Milano, Italy}

 \emailAdd{stefanos.katmadas [at] unimib.it} \emailAdd{alessandro.tomasiello [at] unimib.it}

\abstract{ We consider the BPS conditions of eleven dimensional supergravity, restricted to an appropriate ansatz for black holes in four non-compact directions. Assuming the internal directions to be described by a circle fibration over a K\"ahler manifold and considering the case where the complex structure moduli are frozen, we recast the resulting flow equations in terms of polyforms on this manifold. The result is a set of equations that are in direct correspondence with those of gauged supergravity models in four dimensions consistent with our simplifying assumptions. In view of this correspondence even for internal manifolds that do not correspond to known consistent truncations, we comment on the possibility of obtaining gauged supergravities from reductions on K\"ahler manifolds. }

\begin{document}

 \maketitle

\section{Introduction}

The research fields of flux compactifications of string theories and of black hole physics in lower dimensions have been cross-fertilized repeatedly. The structure of BPS black holes near their horizons, the so-called attractor region, has inspired the search for flux vacua, while the nontrivial features of flux vacua have been useful in extending the toolkit of relevant Ans\"atze for black holes in the lower dimensional compactifications. 


Recently, the understanding of BPS black hole solutions in AdS spacetimes has progressed considerably, so that the connection to higher dimensional compactifications can be explored. In four spacetime dimensions, BPS black hole solutions in gauged supergravity have been obtained for Fayet--Iliopoulos gaugings, starting with the work of \cite{cacciatori-klemm-mansi-zorzan,cacciatori-klemm}, which showed the existence of regular spherically symmetric BPS black holes. Subsequent extensions uncovered  fully analytic solutions for symmetric models \cite{dallagata-gnecchi,hristov-vandoren,Barisch:2011ui, katmadas, halmagyi-bh}. 

For theories resulting from string and M-theory reductions, one generally has to extend the scope to include hypermultiplets. Examples of black hole solutions including flows for hypermultiplets have been discussed in the framework of consistent reductions of M-theory on coset spaces to four dimensional gauged supergravity \cite{cassani-koerber-varela}, resulting in regular numerical solutions \cite{halmagyi-petrini-zaffaroni-bh, Erbin:2014hsa}. These examples are particularly interesting because the consistent reduction allows for a lift of the solutions to M-theory, in order to obtain solutions to the eleven dimensional theory.

In this paper, we consider a more general framework, exploring asymptotically AdS$_4$ black holes in M-theory, assuming internal Sasaki--Einstein seven-dimensional manifolds $M_7$ which are not cosets and which are \emph{regular}, meaning that they can be written as a circle fibration over a six dimensional K\"ahler--Einstein base space $M_6$. A static black hole solution corresponds to a continuous deformation of this Sasaki--Einstein manifold along a radial direction, terminating at the black hole horizon, where an attractor solution with enhanced symmetry arises \cite{Gutowski:2011xx,Gutowski:2012eq}. Our starting point to obtain the relevant flow equations is the classification of BPS solutions in M-theory in \citep{gauntlett-pakis}, which we use to define an appropriate ansatz for static, asymptotically AdS$_4$, black holes preserving two supercharges.\footnote{One might also have considered using the formalism of \cite{rt} to find solutions in type II theories rather than in M-theory. The supercharges of AdS black holes, however, are not immediately compatible with the structure considered in \cite{rt}; one would need to extend it by doubling the amount of internal spinors one considers.}

The result is a set of flow equations that are formally identical with the known flow equations for gauged supergravity models arising from M-theory reductions. More specifically, in the case of symmetric models, the four-dimensional flow equations of \cite{halmagyi-petrini-zaffaroni-bh} can be cast in a form involving the $I_4$ quartic invariant, following \cite{katmadas}. The equations we find in this paper have exactly the same form, but with $I_4$ replaced by the Hitchin functional \cite{hitchin-gcy} on $M_6$; the main equation is given in (\ref{eq:fullflow}) below. This result gives an M-theory explanation of the reformulation in \cite{katmadas}, and shows that it is valid for non-symmetric models as well.\footnote{Progress in non-symmetric models was also achieved recently in \cite{klemm-marrani-petri-santoli}.}

While we are not aware of any reductions of M-theory to four dimensional supergravity on general K\"ahler--Einstein base spaces, the form of the flow equations we find makes it tantalising to conjecture that such reductions might indeed be possible to carry out in more general situations than cosets. This might be important also in view of the recent mathematical progress in finding such spaces: a stability condition was recently proved \cite{chen-donaldson-sun}, which has already been yielding concrete results \cite{datar-szekelyhidi}.

This paper is organised as follows. In section \ref{sec:flow-gen} we give our general strategy for obtaining black hole solutions from M-theory, specified to an ansatz for static solutions that only depend on the radial variable. We impose that ansatz to obtain a set of flow equations for the radial evolution of fields in terms of conserved charges of the eleven dimensional theory. We then proceed in section \ref{sec:4d} to further specify these flow equations to the case of asymptotically AdS$_4$ solutions, by changing to variables that naturally appear in four dimensions. This is made systematic by the use of polyform language and of the Hitchin functional on the K\"ahler base space, which we use to define appropriate operators that appear in four dimensional theories. Finally, section \ref{sec:4dsugra} is devoted to a short overview of the BPS flow equations for static black holes in gauged supergravity and the comparison with the flow equations obtained from the M-theory reduction. Given that the match extends beyond the cases connected to coset spaces, we discuss various possibilities and future directions towards connecting more general gauged supergravity models to M-theory reductions on K\"ahler spaces.

\section{Black hole flow equations from eleven dimensions}
\label{sec:flow-gen}

In this section, we consider static backgrounds of eleven dimensional supergravity on a six dimensional K\"ahler manifold times a circle, assuming that two supercharges are preserved. We start by giving a short review of the M-theory BPS backgrounds of \cite{gauntlett-pakis}, which preserve an SU(5) structure and generically allow for a single supercharge. We then spell out our ansatz to obtain static black hole backgrounds, which we then implement to obtain flow equations for the moduli that interpolate between AdS$_2\times$S$^2$ and AdS$_4$.

\subsection{BPS solutions of eleven-dimensional supergravity}

The bosonic fields of $D=11$ supergravity consist of a metric, $g$, and a three-form potential $A$ with four-form field strength $F=dA$. The action for the bosonic fields is given by 
\begin{eqnarray}
S=\frac{1}{2\kappa^2}\int d^{11} x {\sqrt{-g}}R
-\frac{1}{2}F\wedge *F - \frac{1}{6}C\wedge F\wedge F \,, 
\end{eqnarray}
where $F=dC$. The equations of motion are thus given by 
\begin{subequations}
\begin{gather}\label{eq:EOM-11D}
	R_{\mu\nu}-\frac{1}{12}(F_{\mu \rho \lambda \sigma}F{_{\nu}}{^{\rho \lambda \sigma}}-
	\frac{1}{12}g_{\mu\nu}F^2) = 0 \,,
	\\
	d*_{\scriptscriptstyle 11}F+\frac{1}{2}F\wedge F = 0 \,.
\end{gather}
\end{subequations}
We are interested in bosonic solutions to the equations of motion that preserve at least one supersymmetry, as described in \cite{gauntlett-pakis}. The presence of a Killing spinor implies the existence of a Killing vector which we will assume to be timelike throughout this paper. The metric can then be written as a time fibration over a ten-dimensional manifold, $M_{10}$, as
\begin{equation}\label{eq:gen-metr}
 ds^2 = - \Delta^2 (dt + \omega)^2 + \Delta^{-1} ds^2{ (M_{10}) }\,.
\end{equation}
Here, $\Delta$ and $\omega$ are a function and a one-form on $M_{10}$, which is assumed to be equipped with an $SU(5)$ structure $(\Jd{5}, \Od{5})$, where $\Jd{5}$ is the symplectic $(1,1)$ form and $\Od{5}$ is the holomorphic $(5,0)$ form. There is a single general constraint on the torsion classes of this $SU(5)$ structure, given by
\begin{equation}\label{eq:tor-con}
 \Re W_5 = -12\,d \ln \Delta \,,
\end{equation}
where the two one-form torsion classes $W_4$ and $W_5$ are defined as
\begin{equation}
 W_4 = \Jd{5} \llcorner d \Jd{5} \,, \qquad W_5 = \Re\Od{5} \llcorner d ( \Re\Od{5} )\,.
\end{equation}
Here, $A \llcorner B$ denotes the standard contraction of the components of an $n$-form, $A$, with the first $n$ indices of an $m$-form, $B$, for $m>n$.
The four-form field strength is fixed in terms of these data as
\begin{align}\label{eq:gen-F}
 F = &\, - d\left[ (dt + \omega)\wedge \Jd{5} \right]  + \Lambda
\CR
    &\, + \frac12\,*d\left( \Delta^{-3/2} \Re \Od{5}\right) - \frac12\,*\left[ \Jd{5}\wedge d\left( \Delta^{-3/2} \Re \Od{5}\right)  \right]\wedge \Jd{5}
\\
    &\, - \frac1{16}\, *\left[ \left( W_5 + 4\, W_4\right)\Delta^{-3/2} \Re W_5  \right] \,. \nonumber
\end{align}
Here, $*$ denotes the Hodge dual on $M_{10}$ and $\Lambda$ is a $(2,2)$ four-form on the base that satisfies the constraint
\begin{equation}\label{eq:Lambda-constr}
\Jd{5} \llcorner \Lambda = 2\, d\omega\,.
\end{equation}
Note that \eqref{eq:Lambda-constr} can be solved by decomposing $\Lambda$ in terms of a primitive $(2,2)$ form and the symplectic form $J_{(5)}$, as in the original derivation of \citep{gauntlett-pakis}. However, we prefer the constraint \eqref{eq:Lambda-constr}, as the relevant Ans\"atze for black hole solutions are naturally given in terms of $\Lambda$.
 
\subsection{Black hole Ansatz}
\label{sub:bh-ansatz}

In order to describe black hole solutions, one must make assumptions on the form of the manifold $M_{10}$ in \eqref{eq:gen-metr}. Here, we are ultimately interested in static, spherically symmetric black hole solutions asymptotic to the product of AdS$_4$ with a regular Sasaki-Einstein manifold. With these assumptions, the solution may only depend on a single, radial, variable, so we assume the manifold $M_{10}$ to be the product of a radial direction $\mathbb{R}_+$, parametrised by a coordinate $r$, and a nine-dimensional circle fibration: 
\begin{equation}\label{eq:M10-top}
 M_{10}= \mathbb{R}_+ \times M_9 \ ,\qquad S^1\hookrightarrow M_9 \to M_8\ .
\end{equation}
Here, $M_8$ is an eight dimensional base manifold and the $S^1$ will ultimately correspond to the circle fibration of the regular Sasaki-Einstein manifold. One may consider various assumptions on the form of the manifold $M_8$ and the circle fibration over it, corresponding to solutions in various spacetime dimensions. For spherically symmetric black hole solutions in AdS$_4$, $M_8$ must have an ${\rm SU}(2)$ isometry and will be taken to be itself a product:
\begin{equation}
	M_8 = S^2\times M_6\ ,
\end{equation}
where $S^2$ is a round sphere. The $S^1$ in \eqref{eq:M10-top} will in general be fibred over both the $S^2$ and the $M_6$. It then follows that $M_9$ can also be thought of as a fibration of $M_7$ over $S^2$, where $M_7$ is the total space of the fibration of the $S^1$ over $M_6$: $S^1\hookrightarrow M_7 \to M_6$. 

At $r\to \infty$, the geometry should be asymptotic to a vacuum solution;\footnote{To be more precise, regular AdS$_4$ black hole solutions such as \cite{cacciatori-klemm} are asymptotic at $r\to \infty$ to a solution with AdS$_4\times M_7$ metric, but where a magnetic flux for the graviphoton is also present: this is a remnant of the magnetic charge that does not die out at infinity. This kind of asymptotics was dubbed ``magnetic AdS'' in \cite{Hristov:2011ye}.} for simplicity in this paper we will achieve this by imposing that $M_7$ should be a Sasaki--Einstein manifold at infinity. $M_6$ then has to be asymptotic at $r\to \infty$ to a K\"ahler--Einstein manifold of positive curvature. Again for simplicity, as we anticipated, we will take $M_6$ to be K\"ahler \emph{along the entire flow}, or in other words for any $r$.

%
%

Assuming dependence on the single radial variable, denoted by $r$, the four-form field strength simplifies as well, since
\begin{equation}
  \Re W_5 = -12\,d \ln \Delta \quad \Leftrightarrow \quad d\left( \Delta^{-3/2} \Re \Od{5}\right) =0 \,.
\end{equation}
It follows that the second line in \eqref{eq:gen-F} vanishes identically. In addition, we note that the spherical symmetry we assumed does not allow the four-form to have a single leg on the sphere, so that we must impose that the third line of \eqref{eq:gen-F} vanishes as well, as
\begin{equation}\label{eq:no-1}
 W_5 + 4\, W_4 = 0 \,.
\end{equation}
The final form for the gauge field field strength reads
\begin{align}\label{eq:red-F}
 F = &\, -d\left[(dt+\omega) \wedge \Jd{5} \right] + \Lambda\,,
\end{align}
where the magnetic component $\Lambda$ is still subject to \eqref{eq:Lambda-constr} above. 
In order to satisfy the Bianchi identity $dF=0$, moreover, $\Lambda$ must be closed.
Note that the assumption of spherical symmetry does not imply that the rotational one-form $\omega$ is identically zero, as it can have a nontrivial component along the internal $S^1$ in \eqref{eq:M10-top}. It is straightforward to consider rotating solutions in AdS$_4$ along similar lines, by assuming all metric components to depend on more than the radial variable and allowing $\omega$ to have components along the sphere.

With this Ansatz, we can be more explicit about the equation of motion for the three-form gauge field, which will be useful in the following. Inserting \eqref{eq:red-F} in the equation of motion for the gauge field in \eqref{eq:EOM-11D}, we find
\begin{align}\label{eq:red-EOM}
 d*_{\scriptscriptstyle 11}F+\frac{1}{2}F\wedge F = 
 &\, d\left[ (dt+\omega) \wedge \left( *(\Lambda - \Jd{5}\wedge d\omega) - \Jd{5}\wedge (\Lambda - \tfrac12\,\Jd{5}\wedge d\omega) \right) \right]
\CR 
 &\, + d\left( \Delta^{-3}*d\Jd{5} \right) + \frac12\,\Lambda\wedge\Lambda\,,
\end{align}
so that each of the two terms in the right hand must vanish separately. It turns out that the constraint \eqref{eq:Lambda-constr} is precisely equivalent to the timelike component under the
derivative in \eqref{eq:red-EOM}, so that the three-form equation of motion reduces to a Poisson equation for $\Jd{5}$ on $M_{10}$, as
\begin{equation}
 d\left( \Delta^{-3}*d\Jd{5} \right) + \frac12\,\Lambda\wedge\Lambda =0\,.
\end{equation}
In order to define an electric charge associated to the three-form, one needs to strip off a derivative, so that a three-form $\lambda$ exists with the property  $\Lambda=d\lambda$, at least locally. One may then define a conserved electric charge by integrating over an appropriate seven-cycle $\Omega_7$:
\begin{equation}\label{eq:Q-ch-gen}
 Q = \int_{\Omega_7}\left( \Delta^{-3}*d\Jd{5} + \frac12\,\lambda\wedge\Lambda \right)\,.
\end{equation}
We will make this more precise later for our class of solutions.

We have now taken care of the supersymmetry equations, of the Bianchi identity $dF=0$ (by requiring $\Lambda$ to be closed), and of the flux equation of motion (\ref{eq:red-EOM}). By \cite{gauntlett-pakis} it now follows that the Einstein equations (\ref{eq:EOM-11D}) are also satisfied.

\subsection{Flow equations}
\label{sub:flow}

We now start imposing the assumptions spelled out above, taking the metric on $M_{10}$ to be given by 
\begin{equation}\label{eq:11d-metr}
ds^2 = - \Delta^2 (dt + \omega)^2 + \ee^{2W} dr^2 +  \ee^{2V} \theta^2 + ds^2 (M_{8}) \,,
\end{equation}
where $\ee^{W}$ and $\ee^{V}$ are functions of $r$ and we reabsorbed the factor of $\Delta$ from all the spatial directions, compared to \eqref{eq:gen-metr}, for simplicity. The one-form, $\theta$, here is assumed to correspond to the $S^1$ fibration over $M_8$ in \eqref{eq:M10-top}, so that it is of the type
\begin{equation}
 \theta = d\psi + A \,,
\end{equation}
where $A$ is a one-form on $M_8$ and $\psi$ is an angular coordinate. In any case, we will not use this parametrisation below. We will assume the manifold $M_8$ to have an SU(4) structure $(\Jd{4}, \Od{4})$, and we will restrict to the case
\begin{equation}\label{eq:M8-structure}
 d \Jd{4} = 0\,, \qquad  d \Od{4} = \mathrm{i}\, E\,\ee^{V} \theta \wedge \Od{4} \,,
\end{equation}
where the real one-form $\theta$ and $\ee^{2V}$ are the quantities appearing in \eqref{eq:11d-metr}.
Similarly, $E$ represents one of the torsion classes on $M_8$.\footnote{One could consider turning on more such classes, but we restrict to this case for simplicity.} When embedded in $M_{10}$ as in \eqref{eq:M10-top}, both the torsion $E$ and $\ee^{2V}$ are in general promoted to real functions of the radial variable parametrizing $\mathbb{R}_+$. However, in this paper we will
only consider K\"ahler deformations, and thus we will assume $E$ to be a constant, since it would only depend on complex structure moduli. The conditions (\ref{eq:M8-structure}) are met in the case that we will be eventually interested in, as anticipated at the beginning of section \ref{sub:bh-ansatz}: $M_8=S^2\times M_6$, with $M_6$ a K\"ahler--Einstein. In that case, $\Jd{4}$ will be factorized in the obvious way; $\Od{4}$ will be of the form $e\wedge e^{i\xi} \Od{3}$, where $e$ is a $(1,0)$ form on the $S^2$, and $\Od{3}$ a $(3,0)$ form on $M_6$. Notice that there are several $e$'s that one can pick on the $S^2$, rotated by an SO(3), corresponding to the fact that our solutions will have two supercharges.

The deformations of the K\"ahler form correspond to vector multiplet moduli in a lower dimensional supergravity truncation, when that exists. These can be defined by expanding on a basis $\{\omega_p\}$ of the $(1,1)$ cohomology on $M_8$, on which the K\"ahler form can be expanded as 
\begin{equation}\label{eq:Kah-exp}
 \Jd{4} = t^p \, \omega_p \,,
\end{equation}
where the $t^p$ are the K\"ahler moduli. We will work directly with the K\"ahler form, without enforcing this expansion, using generic identities such as
\begin{align}\label{eq:Kah-metr-4}
 *_{4}\,\omega = &\, \frac{1}{2} \Jd{4}\wedge \Jd{4}\wedge \omega 
     - \left(\Jd{4} \llcorner \omega \right)\, \frac{1}{3!} \Jd{4}\wedge \Jd{4}\wedge \Jd{4} \,,
\end{align}
for any $(1,1)$ form $\omega$. We also define the volume of $M_8$ as
\begin{equation}
 V_4 = \int \frac{1}{4!} [\Jd{4}]^4 = \int \Od{4} \wedge \bOd{4}\,.
\end{equation}

In terms of these objects, the SU(5) structure on $M_{10}$ in \eqref{eq:11d-metr} is given by the forms
\begin{equation}
 \Jd{5} = \Delta\,\ee^{W+V} dr \wedge \theta + \Delta\,\Jd{4}\,,
\qquad
\Od{5} = \Delta^{5/2}\,( \ee^{W} dr + \mathrm{i}\, \ee^{V} \theta )\wedge \Od{4}\,,
\end{equation}
which satisfy the defining condition $\Jd{5} \wedge \Od{5} = 0$, as well as
\begin{equation}\label{eq:structure-ders}
\begin{split}
	 \qquad d\Jd{5}=&\, \left( \partial_r (\Delta\,\Jd{4}) -\Delta\,\ee^{W+V} d\theta \right)\wedge dr \,, 
	\\
	d\Od{5}=&\, \left( \partial_r \ln \left( \Delta^{5/2} \sqrt{V_4} \ee^V \right) -E\,\ee^{W-V} \right)\, dr\wedge \Od{5} \,.	
\end{split}
\end{equation}
In deriving this, we assumed that all complex structure moduli on $M_8$ are frozen, so
that the derivative of a $(p,q)$-form on $M_8$ is again a $(p,q)$-form.
The relevant torsion classes then read
\begin{equation}
\begin{split}
	 W_4 = &\, \Jd{4} \llcorner \left( \Delta^{-1} \partial_r(\Delta\,\Jd{4}) -\ee^{W+V} d\theta \right) \,dr \,,
	\\
	 W_5 = &\,-8\,\left( \partial_r \ln \left( \Delta^{5/2} \sqrt{V_4} \ee^V \right) -E\,\ee^{W-V} \right)\,dr \,.	
\end{split}
\end{equation}
This can be used in \eqref{eq:tor-con} and \eqref{eq:no-1}, to obtain
\begin{align}\label{eq:sing-flow}
	 \partial_r \ln \left( \Delta\,\sqrt{V_4} \ee^V \right) = &\, E\,\ee^{W-V} \,, 
	\CR
	\Jd{4} \llcorner \partial_r \Jd{4} -\ee^{W+V} \Jd{4} \llcorner d\theta = &\, - \Delta^{-1}  \partial_r\Delta \,,	
\end{align}
respectively. 

We now turn to the remaining objects in \eqref{eq:red-F}, namely the rotational one-form $\omega$ and the four-form $\Lambda$. Since we are interested in solutions that appear static from a four-dimensional point of view, $\omega$ may not have any components along those directions, but we will allow for a nontrivial component along the circle parametrized by $\theta$, so as to obtain a nontrivial charge for the associated Kaluza--Klein gauge field. We therefore take
\begin{equation}\label{eq:omega-def}
\omega = M\, \theta \,,
\end{equation}
where $M$ is a function of the radial variable only. Similarly, we adopt the following Ansatz for the four-form $\Lambda$:
\begin{equation}\label{eq:Lambda-def}
\Lambda = d \left[ C \wedge \theta \right] + \Pi \ ,\qquad
        C \equiv B + M\,\Delta\,\Jd{4} 
\end{equation}
where $B$ is a $(1,1)$ form and $\Pi$ is a constant $(2,2)$ form flux, both defined on $M_8$. Note that the first term in \eqref{eq:Lambda-def} is chosen so that $B$ can be interpreted as the $B$-field of Type IIA string theory; its components upon expansion on a basis as in \eqref{eq:Kah-exp} are identified with the vector multiplet axions from a four-dimensional point of view \cite{cassani-koerber-varela}. Possible hyper-scalars in four dimensions would be described by adding $(3,1)$ forms in the total derivative in \eqref{eq:Lambda-def}, but we have set these to zero in this paper. Furthermore, we require the condition 
\begin{equation}\label{eq:flux-con}
\Pi\wedge\Pi=0\,,
\end{equation} 
in order to ensure that there is a local expression for the electric charges defined in \eqref{eq:Q-ch-gen}, which demands that $\Pi\wedge\Pi$ be trivial in cohomology.

More explicitly, let us consider the electric charges \eqref{eq:Q-ch-gen} carried by the field strength along the various seven-cycles $\Omega^p$. These read
\begin{align}\label{eq:charge-def}
 Q_p  \equiv &\,
 \frac1{N_p}\, \int_{\Omega^p}\! \left( * F + \frac12\,A\wedge F \right)
\\
= &\, 
 \frac1{N_p}\, \int_{\Omega^p}\! \left(
\Delta^{-1} \ee^{V-W}*_4\left( \partial_r(\Delta\,\Jd{4}) -\Delta\,\ee^{W+V} d\theta \right) 
+ \frac{1}{2}\, C \wedge C \wedge d\theta + C\wedge\Pi
\right) \wedge \theta
\nonumber \,,
\end{align}
where the normalisation constant $N_p$ is the volume of $\Omega^p$ for trivial moduli.
Note that this definition would indeed be impossible without the condition \eqref{eq:flux-con}, as one would not be able to write the eight-form $F\wedge F$ as a total derivative.

The definition \eqref{eq:charge-def} can be viewed as a first order flow equation for the K\"ahler form; it can be recast as
\begin{align}\label{eq:charge-fin}
Q = &\, \ee^{V-W} \, \partial_r\left( \tfrac1{3!}\, \Jd{4} \wedge \Jd{4}\wedge \Jd{4} \right) 
-\ee^{2V}\tfrac1{2}\, \Jd{4} \wedge \Jd{4}\wedge d\theta
+ \frac{1}{2}\, C \wedge C \wedge d\theta + C\wedge\Pi
\,.
\end{align}
We used the second of \eqref{eq:sing-flow} and the identity \eqref{eq:Kah-metr-4}. For convenience we have combined the electric charges into a six-form $Q$, with the understanding that the actual charges are the components of this form along an appropriate basis on $M_8$, following \eqref{eq:charge-def}:
\begin{equation}
Q_p = \frac1{N_p}\, \int_{\Omega^p} Q \wedge \theta\,.
\end{equation}
From the three-form equation of motion and the definition \eqref{eq:charge-def}, one finds the important constraint
\begin{equation}\label{eq:ch-con}
 d\theta\wedge Q \simeq 0 \,.
\end{equation}
$\simeq$ stands for cohomological equality, so that the integral of the left hand side of \eqref{eq:ch-con} vanishes upon integration over $M_8$.

The evolution of the two-form $B$ is described by a flow equation obtained by inserting \eqref{eq:Lambda-def} into the constraint \eqref{eq:Lambda-constr}. When written in components along $M_8$ and along $dr\wedge\theta$, one finds
\begin{gather}
\Delta^{-1} \ee^{-V-W} \partial_r C + \Delta^{-1} \Jd{4}\llcorner(C\wedge d\theta + \Pi)
= 2\, M \, d\theta\,,
\CR
\Delta^{-1} \Jd{4}\llcorner (\partial_r C) =  2\,\partial_r M\,.
\label{eq:B-flow-def}
\end{gather}
We used the definition \eqref{eq:omega-def} to compute the right hand side.

There is a final flow equation, corresponding to the conserved angular momentum along the U(1) isometry $\xi$ dual to $\theta$. This is naturally computed by the (matter modified) Komar integral associated to $\xi$, the so called Noether potential. In Appendix \ref{app:noe-pot} we give a short discussion of the steps required to define this conserved integral. A bottom-up approach is explained in some detail in \cite{Hanaki:2007mb} for the closely related case of five-dimensional supergravity, which contains ordinary gauge fields instead of the three-form. However, the same steps can be followed to obtain a Komar integral, which can be written as 
\begin{align}\label{eq:J-def}
 q_0  \equiv &\,
 \frac1{N_9}\, \int_{M_9}\! 
 \left( * d\xi + (\xi\cdot A) \wedge * F + \tfrac13\, (\xi\cdot A) \wedge A \wedge F \right) 
\CR
= &\, 
 \frac1{N_9}\, \int_{M_9}\! \left[
\Delta^{-1} \ee^{3V-W}\,\tfrac1{4!}[\Jd{4}]^4\,\partial_r(\Delta^{2} \ee^{-2V} \, M)
\right.
\CR
&\, \qquad \qquad
+( C - M\,\Delta\,\Jd{4} )\wedge \left( \ee^{V-W} \, \partial_r\left( \tfrac1{3!}\, \Jd{4} \wedge \Jd{4}\wedge \Jd{4} \right) 
-\ee^{2V}\tfrac1{2}\, \Jd{4} \wedge \Jd{4}\wedge d\theta \right)
\CR
&\, \qquad \qquad
\left.+ \tfrac{2}{3}\,C \wedge C \wedge\left(\tfrac{1}{2}\, C \wedge d\theta + \Pi\right)
\right] \wedge \theta
 \,,
\end{align}
where the normalisation constant $N_9$ is the volume of $M_8\times S^1$ for trivial moduli.
The interested reader can find more details on the general definition of conserved charges dual to Killing vectors in \cite{wald-entropy} and references therein.

The flow equations (\ref{eq:sing-flow}), \eqref{eq:charge-fin}, (\ref{eq:B-flow-def}), \eqref{eq:J-def} for the K\"ahler moduli and the $B$ field describe the full flow of the solution. In the following, we proceed to recast the same set of equations in a form that is more suggestive from the four-dimensional point of view.

\section{Four-dimensional black holes} 
\label{sec:4d}

In the previous discussion, we emphasized the properties of the eight-dimensional compact manifold $M_8$, so that the four-dimensional interpretation is somewhat obscured. Indeed, by choosing appropriate Ans\"atze for $M_8$ one may hope to describe solutions in various spacetime dimensions. In this section, we will focus on black holes in four dimensions, taking $M_8=S^2\times M_6$, with the $S^2$ having the role of the space surrounding the black hole, as was already anticipated by our choice of Ansatz explained around \eqref{eq:M8-structure}. We then recast the flow equations in terms of variables most natural for a gauged supergravity in four dimensions, using the language of polyforms defined on $M_6$.

\subsection{Four-dimensional Ansatz} 
\label{sub:4d}

A convenient metric Ansatz for a four-dimensional solution is inspired by the special case of the consistent reductions described in \cite{cassani-koerber-varela}, given by
\begin{align}\label{eq:metr-2}
ds_{11}^2= &\, \ee^{2V}\gamma^2 \ee^{-K} \left( - \gamma^2 \ee^{2U} (dt + M\,\theta)^2 + \ee^{-2U} dr^2 + \ee^{2\chi} ds^2(S^2) \right) 
 + \ee^{-V} ds^2(M_6)+ \ee^{2V} \theta^2
 \CR
 = &\, \ee^{2V}\gamma^2 \ee^{-K} ds^2_4 
 + \ee^{-V} ds^2(M_6)+ \ee^{2V} \left( \theta - \gamma^2 \ee^{2U} M\,dt\right)^2
\,,
\end{align}
where in the second line we rewrote the metric in a way that exhibits the asymptotically AdS$_4$ static metric, given by
\begin{equation}
 ds^2_4 = - \ee^{2U} dt^2 + \ee^{-2U} dr^2 + \ee^{2\chi} ds^2(S^2) \,.
\end{equation}
The functions $V$, $U$, $\chi$ and $M$ and the metric on $M_6$ all depend on $r$ only.  $\gamma$ is related to $M$ by
\begin{equation}\label{eq:gamma-def}
M = -  \ee^{-U} \ee^{-K/2} \gamma^{-2}\sqrt{1-\gamma^2}\,,
\end{equation}
so that $\gamma=1$ corresponds to a static metric in eleven dimensions. Here, $\ee^{-K}$ is the K\"ahler potential defined as
\begin{equation}\label{eq:K-vl-def}
 \vl_3 = \frac{1}{3!}\,\jg \wedge \jg \wedge \jg \,,
 \qquad
 \ee^{-K} = \int_{M_6}\!\vl_3 \,,
\end{equation}
where $\jg $ is the K\"ahler form of the base $M_6$. The dilaton in four dimensions is defined by 
\begin{equation}
 \ee^{2\phi/3} = \ee^{V} \ee^{K/3}\gamma \,,
\end{equation}
while the remaining metric functions can be determined by comparing with \eqref{eq:11d-metr}, to find
\begin{align}\label{eq:ch-var-4d}
 \Delta =&\,  \ee^{2\phi/3} \ee^{U} \ee^{K/6}\gamma \,, 
 \nonumber\\
\ee^W =&\, \ee^{2\phi/3} \ee^{-U} \ee^{K/6}\,,
 \nonumber\\
\Jd{4} =&\, \ee^{4\phi/3} \ee^{2\chi} \ee^{K/3} \vl_{S^2}+\ee^{-2\phi/3} \ee^{K/3} \jg  \,,
 \nonumber\\
V_4 =&\,\ee^{-2\phi/3} \ee^{2\chi} \ee^{K/3}\,.
\end{align}
Here, $\vl_{S^2}$ is the volume of the $S^2$.  
Note that at this point these equalities define a change of variables rather than a further refinement of our Ansatz.

As we have a four-dimensional flow in mind, one of the charges will be special: the charge in \eqref{eq:charge-fin} that corresponds to a flux over $M_6\times {\rm U}(1)$, which is to be viewed as the internal space. This particular charge
is the Freud--Rubin parameter of the AdS$_4$ compactification, usually denoted as $e_0$. We therefore decompose $Q$ in \eqref{eq:charge-fin} in a four-form, $\qg$, and a six-form on $M_6$:
\begin{align}
Q = -\qg\wedge \vl_{S^2} + e_0\,\ee^{-K} \vl_3\,;
\end{align}
we used the definitions in \eqref{eq:K-vl-def}. Similarly, we decompose each of $C$ and $d\theta$ in terms of forms on $M_6$, as
\begin{align}\label{eq:dec-4d}
C =&\, c^0\, \vl_{S^2} + \cg 
=  \left( M\,\ee^{2\phi}\ee^{2\chi}\,\ee^{U} \gamma\,\ee^{K/2} \right)\, \vl_{S^2} + \bg + M\, \ee^{U} \ee^{K/2}\gamma\, \jg \,,
 \nonumber\\
d\theta =&\, p^0\, \vl_{S^2} + \mg \,,
\end{align}
where $\cg$ and $\mg$ are two-forms and $c^0$ and $p^0$ are zero-forms. Note that we also gave the explicit expressions for $c^0$ and $\cg$ in terms of the four-dimensional axions $\bg$, as in \eqref{eq:Lambda-def}, assuming that the component of the $B$-field on the sphere vanishes, consistently with our Ansatz for the four-dimensional fields. Finally, we choose the flux $\Pi$ to be
\begin{equation}\label{eq:Pi-choice}
\Pi = - \pg\wedge \vl_{S^2}\,.
\end{equation}
where $\pg$ is a $(1,1)$ form on $M_6$, so that \eqref{eq:Pi-choice} solves \eqref{eq:flux-con}.

Using these definitions in \eqref{eq:charge-fin}, one obtains the flow equations
\begin{subequations}\label{eq:ch-4d}
 \begin{align}
e_0\,\ee^{-K} \vl_3 = &\, \ee^{U}\ee^{K/2} \gamma^{-1}\, \partial_r\left(\ee^{-2\phi}\right)\, \vl_3
-\tfrac1{2}\, \gamma^{-2}\jg  \wedge \jg \wedge \mg 
+ \tfrac{1}{2}\, \cg \wedge \cg \wedge \mg 
\label{eq:e0-4d}
 \\
-\qg = &\, \ee^{U}\ee^{-K/2} \, \partial_r\left( \ee^{2\chi} \ee^{-K}\tfrac1{2}\, \jg  \wedge \jg  \right) 
-p^0\,\tfrac1{2}\, \jg  \wedge \jg 
-\ee^{2\phi}\ee^{2\chi}\, \jg \wedge \mg 
\CR
&\, + \cg\wedge \left(\tfrac12\, p^0\,\cg + c^0\,\mg  -\pg \right) \,.
\label{eq:q-4d}
 \end{align}
\end{subequations}
These must be supplemented by the two scalar equations in \eqref{eq:sing-flow}, which read 
\begin{equation}\label{eq:sing-flow-4d}
\begin{split}
	\partial_r \left( \ee^{U} \ee^{\phi} \ee^{\chi} \right) =&\, E\,\ee^{\phi} \ee^{\chi}\ee^{K/2} \gamma \,, 
	\\
	\partial_r \left( \ee^{U} \ee^{K/2} \ee^{2\chi} \gamma\right) =&\, \gamma^{-1} p^0 + \gamma^{-1}\ee^{2\phi}\ee^{2\chi} (\jg  \llcorner \mg  ) \,.	
\end{split} 
\end{equation}

It is now straightforward to manipulate \eqref{eq:e0-4d} and \eqref{eq:sing-flow-4d} into a form that can be viewed as a scalar
flow in a four-dimensional supergravity theory.
\begin{subequations}\label{eq:ds0-flow}
\begin{gather}
 \partial_r \ee^{-2\phi}  
 =
    \ee^{-U} \ee^{K/2} \gamma\, *_6\left[ 
    e_0\,\vl_3 
    + \tfrac1{2}\, \ee^{-K}\,\left( \gamma^{-2}\jg  \wedge \jg  -\cg \wedge \cg \right)\wedge \mg \right]\,,
\label{eq:dil-flow} \\
 \partial_r \left( \ee^{\chi+U} \right)
 =
 \tfrac12\, \ee^{\chi} \ee^{K/2} \gamma *_6 \left[ ( e_0\ee^{2\phi} + 2\,E )\,\vl_3 
 + \tfrac12\,\ee^{-K}\,\ee^{2\phi} \left( \gamma^{-2}\jg  \wedge \jg  - \cg \wedge \cg\right)\wedge \mg  
 \right]\,,
\label{eq:scal-flow} \\
\ee^{\chi+U} \partial_r \left( \ee^{\chi} \ee^{K/2} \gamma \right) 
 = I_{0}  + p^0\,, \label{eq:0-form}
\end{gather} 
\end{subequations}
where we defined the shorthand
\begin{equation}\label{eq:I40-def}
I_{0}=\frac12\, \ee^{2\chi} \ee^{-K} *_6\left( -\gamma^{2}( \ee^{2\phi} e_0 + 2\,E )\,\vl_3  
 + \tfrac12\,\ee^{-K}\,\ee^{2\phi} \left( \jg  \wedge \jg  + \gamma^{2} \cg \wedge \cg\right)\wedge \mg   ) \right) \,,
\end{equation}
which will be useful in the following.

We now turn to the axionic flow equation \eqref{eq:B-flow-def}, starting from the components along $M_6$, which
can be recast as
\begin{equation}\label{eq:2-form}
 \ee^{\chi+U} \partial_r \left( \ee^{\chi} \ee^{K/2} \gamma\, \cg \right)  
= 
 \ee^{2\chi}\ee^{2\phi} \gamma\, \ee^U\ee^{K/2} M\,\mg 
- \ee^{2\chi}\ee^{2\phi} \jg \llcorner(\cg\wedge d\theta) + I_{0}\,\cg
 + \pg\,,
\end{equation}
where we used the definitions \eqref{eq:ch-var-4d}, \eqref{eq:dec-4d} and \eqref{eq:0-form}. Similarly,
the remaining components of \eqref{eq:B-flow-def} along the sphere and $dr\wedge\theta$ can be recast in the form
\begin{subequations}\label{eq:axion-con}
\begin{align}
\left(\ee^U \ee^{-K/2} \gamma\,\partial_rM - \right.&\left. p^0 \ee^{-2\chi} \ee^{-K}\,M\right)\,\vl_3 = 
 \nonumber\\
&\, \tfrac12\,\ee^{-K}\,\jg  \wedge \jg  \wedge \left( d\cg 
 -\ee^{2\phi} M\,\mg  
           + \ee^{-U} \ee^{K/2} \gamma^{-1}\ee^{2\phi} \jg \llcorner(\cg\wedge \mg )\right)
 \label{eq:axion-con-1} \,\\
 M\, \partial_r \ee^{-2\phi}  
 = &\,
    \ee^{-U}\ee^{-K}\, \gamma^{-1}\, *_6\left( \ee^{K/2} M  \,\jg  \wedge \jg \wedge \mg - \ee^{-U} \gamma^{-1} \cg \wedge \jg  \wedge \mg  \right)\,,
    \label{eq:axion-con-2}
\end{align} 
\end{subequations}
to which we will return in due course.

Using the axionic flow equation \eqref{eq:2-form}, it is now straightforward to rewrite the definition of $q$ in \eqref{eq:q-4d} as a flow equation for the K\"ahler moduli in the form
\begin{align}\label{eq:4-form}
&\, \ee^{\chi+U} \partial_r\left(\ee^{\chi} \ee^{K/2} \left( \tfrac1{2}\gamma\, \cg \wedge \cg -\tfrac1{2}\gamma^{-1} \, \jg  \wedge \jg  \right) \right) = 
 \nonumber\\
&\, \qquad\qquad \qquad\qquad \qquad\qquad
\qg + I_{0} \left( \jg  \wedge \jg  + \gamma^{2} \cg \wedge \cg\right)
  - \ee^{2\phi}\ee^{2\chi}\,\gamma^{-2} \jg \wedge \mg 
 \\
&\, \qquad\qquad \qquad\qquad \qquad\qquad \nonumber
+ \ee^{2\phi}\ee^{2\chi}\,\left( 2\,\gamma\, \ee^U \ee^{K/2} M \mg  
  - \jg  \llcorner ( \mg \wedge \cg ) \right) \wedge \cg\,.
\end{align}
This flow equation can be thought of as determining the behaviour of the K\"ahler moduli residing in $\jg $, given that the
axions are similarly determined by the flow equation \eqref{eq:2-form}.

Finally, we turn to the conserved charge \eqref{eq:J-def}, which can also be recast using the Ansatz adapted to four dimensions as
\begin{align}\label{eq:6-form}
\ee^{\chi+U} \partial_r\left(\ee^{\chi}\ee^{K/2} f_0 \right) = &\, 
q_0 
- \left( I_{0} -\tfrac12\,\ee^{2\phi} \ee^{2\chi} \ee^{-K} *_6\left( \jg  \wedge \jg  \wedge \mg  \right) \right) \gamma^{-1} f_0
 \\
&\,
 + \ee^{2\chi} \ee^{K/2}\gamma\,( \ee^{2\phi} e_0 + 2\,E )\,
\left( \ee^{U} M +\ee^{-K/2} \tfrac1{6}\gamma\, *_6( \cg \wedge \cg\wedge \cg) \right) \,,
\nonumber
\end{align}
where we used the shorthand
\begin{equation}\label{eq:F0-def}
 f_0 = 2\,\ee^{U} \ee^{-K/2} M
+ \ee^{-K}\,*_6\left( \tfrac1{6}\gamma\, \cg \wedge \cg\wedge \cg -\tfrac1{2}\gamma^{-1} \, \jg  \wedge \jg \wedge \cg \right)\,.
\end{equation}

This completes the relevant flow equations. However, there are still global constraints, some of which we already alluded to above. For example, the flow must satisfy the constraint \eqref{eq:ch-con}, which upon decomposition on $M_6$ reads as
\begin{equation}\label{eq:ch-con-1}
p^0 e_0\,\ee^{-K} \vl_3 -\qg\wedge \mg =0 \,.
\end{equation}
An additional constraint arises from the requirement \eqref{eq:M8-structure} on the complex structure $\Od{4}$, in the special case where a fibration over a sphere is involved. In particular, this leads to a fixed Chern class of the U(1) fibration, which translates to the condition
\begin{equation}\label{eq:ch-con-2}
 E\,p^0 = n \in \mathbb{Z}\,.
\end{equation}
Here, we conventionally take $n=1$ for a sphere, while negative $n$ corresponds to the case of hyperbolic horizon, which is also allowed and can be treated in exactly the same way, without modifying the flow equations above.

As a summary, in this section we have specialized the flow equations found in section 
\ref{sub:flow} to the case of $M_8=S^2\times M_6$; this resulted in equations (\ref{eq:ds0-flow})--(\ref{eq:6-form}), together with the constraints \eqref{eq:ch-con-1}--\eqref{eq:ch-con-2}. In section \ref{sub:poly}, we will recast these equations in polyform language, which will make them much more compact.

\subsection{Hitchin functional} 
\label{sub:hitchin}

In order to cast the set of flow equations found above in a more systematic way, we find it useful to work with polyforms, namely with formal sums of forms of different dimensions. (From now on, all our forms will be polyforms, and for that reason we will just drop the prefix ``poly''.) In particular, these will allow us to use the language of pure spinors and of generalized complex structures \cite{hitchin-gcy,gualtieri}. These have proven most useful when dealing with the complexities of having two different spinors in the internal six-dimensional space $M_6$ in flux compactifications (see for example \cite{gmpt2,gmpt3}). In this paper we will need a relatively simpler instance of those techniques; we give in this section a lightning review of the main ideas.

We will focus especially on the definition of the Hitchin functional $I_4$, which turns out to provide a natural structure to express the flow equations. As has emerged already in previous work \cite{grana-louis-waldram,hmt}, the Hitchin functional plays a role very similar to the so called quartic invariant of $\cN=2$ supergravity coupled to vector multiplets with a symmetric scalar manifold. 

The first thing to notice is that differential forms on $M_6$ are a representation of a Clifford algebra of signature $(6,6)$. The ``gamma matrices'' are given by the operators  
\begin{equation}
	\Gamma^A \equiv  \{ \del_m \llcorner , dx^m\wedge \}_{m=1,\ldots,6} \ .
\end{equation}
Since $\{ \del_m\llcorner,\del_n \llcorner\}=0=\{ dx^m\wedge,dx^n\wedge\}$ and $\{ \del_m\llcorner,dx^n\wedge\}=\delta_m^n$, the $\Gamma^A$ satisfy indeed a Clifford algebra, with respect to the metric
\begin{equation}\label{eq:Imet}
	\cal I \equiv \left(\begin{array}{cc}0&1_6\\ 1_6&0\end{array}\right)\ .
\end{equation}
Since the $\del_m$ are (pointwise) a basis for the tangent bundle $T$, and the $dx^m$ are a basis for the cotangent bundle $T^*$, one can think of ${\cal I}$ as a metric on $T \oplus T^*$. 

Thus a form on $M_6$ can be thought of as a spinor, in the sense that it is acted upon by the gamma matrices $\Gamma^A$. We can then apply to it the general theory of spinors in diverse dimensions. A \emph{pure spinor} $\phi$ is a form whose annihilator ${\rm Ann}(\phi)\subset T\oplus T^*$ is of dimension 6 --- in other words, there are six linear combinations of the $\Gamma^A$ that annihilate $\phi$. One usually also requires $\phi$ to have non-zero norm
\begin{equation}\label{eq:posnorm}
	(\phi,\bar \phi)\neq 0\ .
\end{equation} 
Here, we used the Mukai pairing, $(\,,)$ of two polyforms, defined as the function such that 
\begin{equation}\label{eq:pairing}
(A\wedge \lambda (B))_6  \equiv (A,B)\, \frac{1}{V_6}\,\vl_6 \ ,
\end{equation}
where ${}_6$ denotes keeping the six-form part only, and on a $k$-form we define $\lambda \omega_k \equiv (-)^{\lfloor \frac k2 \rfloor} \omega_k$; $\vl_6$ denotes the volume form on the manifold and $V_6$ its integral, so that \eqref{eq:pairing} is independent of the volume. Note that \eqref{eq:pairing} is antisymmetric (in six dimensions), while \eqref{eq:posnorm} is required in order for ${\cal J}_\phi$ to be hermitian with respect to the metric ${\cal I}$ we introduced earlier.

To a pure spinor, one can associate in a natural way an almost complex structure ${\cal J}_\phi$ (namely a notion of ``holomorphic index'') on $T \oplus T^*$, essentially by declaring ${\rm Ann}(\phi)$ to be the ``holomorphic'' subspace. ${\cal J}_\phi$ is also called a \emph{generalized} almost complex structure. At every point on $M_6$ it can be viewed as a 12$\times$12 matrix (since $T \oplus T^*$ has dimension 12) which squares to $-1_{12}$. It has a block structure: 
\begin{equation}\label{eq:Jblock}
	{\cal J}^A_B= \left(\begin{array}{cc}
		A^m{}_n & B^{mn}\\ C_{mn} & D_m{}^n
	\end{array}\right) 
\end{equation}
The requirement that it should square to $-1_{12}$ translates into four algebraic identities involving the tensors $A$, $B$, $C$, $D$. Pointwise on $M_6$, the correspondence with the pure spinors can be inverted: namely, to a ${\cal J}$ on $T \oplus T^*$ that squares to $-1_{12}$, one can associate point by point a pure spinor $\phi$. 

A famous example of pure spinor is a $(3,0)$ form $\Omega$, when it exists; this is annihilated by wedging with holomorphic one-forms $dz^i$ and contracting with anti-holomorphic vectors $\del_{\bar i}\llcorner$. Another example, which will be more relevant for us, is the formal exponential $\phi=e^{i\jg}\equiv 1 + i \jg -\frac12 \jg^2 -\frac i6 \jg^3$ on $M_6$. This is annihilated by the six operators of the form $\del_m\llcorner - i \jg_{mn}dx^n\wedge$. In this case, the generalized almost complex structure is, in the language of (\ref{eq:Jblock}), ${\cal J}_\phi= \left(\begin{smallmatrix} 0 & -\jg^{-1}\\ \jg &  0 \end{smallmatrix}\right)$. More generally, $\phi=e^{\bg+i\jg}$, with $\bg$ any real two-form, is also pure. In this case, 
\begin{equation}
	{\cal J}_\phi= 
	\left(\begin{array}{cc}
		1 & 0 \\ b \wedge 0 
	\end{array}\right)
	\left(\begin{array}{cc}
			0 & - \jg^{-1}\\ \jg & 0 
	\end{array}\right)
	\left(\begin{array}{cc}
		1 & 0 \\ -b \wedge 0 
	\end{array}\right)\ .
\end{equation}

Our definitions so far make sense in every even dimensions. In six dimensions, we have a nice characterization of pure spinors. Consider any \emph{real} even form $\rho$ on $M_6$, and consider the 12$\times$12 matrix
\begin{equation}\label{eq:Q}
	{\cal Q}_{AB} \equiv (\rho, \Gamma_{AB} \rho) 
\end{equation}
We define\footnote{Indices here are raised and lowered using the metric ${\cal I}$ in (\ref{eq:Imet}).} 
\begin{equation}\label{eq:Hit}
	I_4(\rho) \equiv -\frac{1}{12}\,{\rm tr}({\cal Q}^2) \equiv \frac{1}{12}\,{\cal Q}_{AB} {\cal Q}^{AB}= \frac{1}{12}\,(\rho, \Gamma_{AB}\rho)(\rho, \Gamma^{AB}\rho)
\end{equation}
Note that we defined the functional $I_4$ with a different overall sign compared to most literature (e.g.~\cite{hitchin-gcy}), for the sake of a more natural connection to supergravity in later sections. If $I_4(\rho)>0$, then
\begin{equation}\label{eq:JQ}
	{\cal J}\equiv \frac{{\cal Q}}{\sqrt{-{\rm tr}({\cal Q}^2)/12}}
\end{equation}
squares to $-1_{12}$: it is a generalized almost complex structure. There should then exist an associated pure spinor. Indeed one can find it, and it has a simple expression:
\begin{equation}\label{eq:phirho}
	\phi= \rho + i \hat \rho \ ,\qquad \hat \rho = -\frac13 {\cal J}\cdot \rho \equiv -\frac16 {\cal J}_{AB} \Gamma^{AB} \rho \ .
\end{equation}
From (\ref{eq:JQ}) we also find $I_4(\rho)= \tfrac14\, (\hat \rho, \rho)^2$.

As an example of how this procedure works, let us consider the polyform $\rho=1+q_4$, with $q_4$ a four-form. In this case, in the language of (\ref{eq:Jblock}), we get $A=D=0$,  $B^{mn}=-\frac2{4!\sqrt{g}}\epsilon^{mnpqrs}q_{pqrs}\equiv -\frac2{\sqrt{g}}\tilde q^{mn}$ (where $\vl_6=\sqrt{g}dx^1\wedge\ldots \wedge dx^6$ in the chosen coordinates) and $C_{mn}= -(\tilde q \llcorner q_4)$. In defining the bivector $\tilde q$ we have never invoked any metric; we have chosen a volume form, which eventually disappears from the final results. Now $I_4$ is proportional to the Pfaffian of $\tilde q$. This eventually produces a $\hat \rho$ such that $\rho+ i\hat \rho$ is a pure spinor. The two-form part of $\hat \rho$ is proportional to $\jg=\frac1{\sqrt{{\rm Pf}(\tilde q)}}\tilde q \llcorner q$. So, for a given four-form $q_4$, we get an explicit way of finding a two-form $\jg$ such that $\jg\wedge\jg = q_4$. In practice, this requires computing all the $4\times 4$ minors of $\tilde q$; for example, in flat indices, $(\tilde q \llcorner q)_{12}=\tilde q^{34}q_{1234}+\ldots= \tilde q^{34} \tilde q^{56}+\ldots$, where the $\ldots$ denote permutations.

Let us now also record some definitions very closely related to $I_4$. We have defined it as a function (of degree 4) of a single form $\rho$, but we can extend it to mean a completely symmetric function of four forms: 
\begin{equation}\label{eq:Hita}
I_4(\alpha_1, \alpha_2, \alpha_3, \alpha_4)\equiv (\alpha_1,\Gamma_{AB} \alpha_2)(\alpha_3, \Gamma^{AB} \alpha_4) + {\rm perm.} \,.
\end{equation}
Likewise, we can define a cubic function $I_4'$ of three forms $\alpha_1$, $\alpha_2$, $\alpha_3$:
\begin{equation}\label{eq:I4'}
	I_4'(\alpha_1,\alpha_2,\alpha_3)\equiv \frac23 \left( (\alpha_1, \Gamma_{AB} \alpha_2) \Gamma^{AB} \alpha_3 + (\alpha_2, \Gamma_{AB} \alpha_3) \Gamma^{AB} \alpha_1 + (\alpha_3, \Gamma_{AB} \alpha_1) \Gamma^{AB} \alpha_2\right)\ .
\end{equation}
Notice that
\begin{equation}
	(\alpha, \Gamma_{AB} \beta)= (\beta, \Gamma_{AB} \alpha) 
\end{equation}
for any forms $\alpha$ and $\beta$; this follows from $(\Gamma_A \alpha, \beta)= -(\alpha, \Gamma_A \beta)$. The function $I_4'$ now takes values in the space of forms.

The normalization in (\ref{eq:I4'}) is such that formally $I_4'(\rho,\rho,\rho)= \delta_\rho (I_4(\rho,\rho,\rho,\rho))$. Recalling (\ref{eq:phirho}) and (\ref{eq:JQ}), this also happens to be
\begin{equation}\label{eq:I4'r}
	I_4'(\rho,\rho,\rho)= 2\,{\cal Q}_{AB} \Gamma^{AB} \rho= - 6\, (\hat \rho, \rho) \hat\rho \ .
\end{equation}
Note that we have stripped off any symmetrisation factors from $I_4(\alpha_1, \alpha_2, \alpha_3, \alpha_4)$, so that whenever two or more arguments coincide, these reappear, so that
\begin{equation}\label{eq:I4fac}
I_4(\alpha_1) \equiv \frac1{4!}\,I_4(\alpha_1, \alpha_1, \alpha_1, \alpha_1)\,.
\end{equation}
For brevity, we also define the shorthand
\begin{equation}\label{eq:I3fac}
I_4'(\alpha_1) \equiv \frac1{3!}\,I_4'(\alpha_1, \alpha_1, \alpha_1)\,.
\end{equation}
Together with \eqref{eq:I4fac} this is the only instance where a single argument appears.

Along these lines we can also define a ``second derivative'' $I_4''(\alpha,\beta)$, which is now a matrix that acts on a form and gives another form. In particular we have
\begin{equation}\label{eq:I4''}
 I_4''(\rho,\rho)= \frac{2}{3}\left( (\rho, \Gamma_{AB} \rho) \Gamma^{AB}+ 2 \Gamma^{AB}\rho (\rho,\Gamma^{AB}\ \cdot)  \right) \ .
\end{equation}
Notice that $I_4''(\rho,\rho) \omega = I_4'(\rho,\rho,\omega)$. 

Having defined the functional $I_4$, we will now argue that it essentially plays the same role as the quartic invariant ${\cal I}_4$ that sometimes appears in four-dimensional supergravity theories. In the context of BPS black holes, there are two contexts in which this quartic invariant plays a role. 

First, in asymptotically Minkowski black holes, the entropy can be written as $\sqrt{{\cal I}_4(\Gamma)}$, where $\Gamma$ is a vector of electric and magnetic charges. This can be reproduced from type II in ten dimensions \cite{hmt}: the story goes roughly as follows. The attractor equation reads schematically \cite{denef-flows,hmt} $f={\rm Re} \phi$, where $f$ is an internal form that collects the charges and $\phi$ is a pure spinor. (When the internal space $M_6$ is a Calabi--Yau, in IIA we have $\phi=e^{i\jg}$, in IIB $\phi=\Omega$.) This equation can be solved by applying (\ref{eq:phirho}) above to $f=\rho$. Notice that this makes sense only if $I_4(f)>0$. This is related to the fact that the black hole entropy is proportional to $\int \sqrt{I_4(f)}$. 

Second, and more relevant for our present purposes, the supergravity BPS equations for asymptotically AdS$_4$ were written out for static backgrounds in \cite{halmagyi-petrini-zaffaroni-bh} and can be reformulated\footnote{Note that in \cite{katmadas} this was done in the special case without hypermultiplets, but the result can be extended trivially to theories with hypermultiplets, as will be shown in section \ref{sec:4dsugra}.} in terms of the quartic invariant ${\cal I}_4$ as in \cite{katmadas}. The formal properties of the quartic invariant ${\cal I}_4$ that were important in \cite{katmadas} were equations (2.10), (B.4)--(B.6) in that paper. Very close analogues of those equations are valid for $I_4$: \cite[Eq.(2.10)]{katmadas} becomes
\begin{subequations}\label{eq:I4prop}
\begin{align}
	\label{eq:I4'J}
	I_4'(\rho, \rho, \omega)&= 2 (\hat\rho, \rho)\,{\cal J}\cdot \omega + 4 (\rho,\omega)\,\rho\\
	\label{eq:I4*}
	&= - 2 (\hat \rho, \rho) * \lambda \omega -4 (\rho,\omega) \rho + 8 (\hat \rho, \omega) \hat \rho
	\ .
\end{align}
\end{subequations}
The first expression, (\ref{eq:I4'J}), is always valid, while (\ref{eq:I4*}) holds if $\omega$ satisfies the property $(\Gamma^A \phi_-, \omega)= (\Gamma^A \bar \phi_-, \omega)=0$, where $\phi_-$ is a pure spinor compatible with $\phi$. (In the ``generalized Hodge diamond'', see e.g.~\cite[Eq.(A.20)]{gmpt3}, this means that $\omega$ belongs to the central column.) For our applications $\phi= e^{i\jg}$, $\phi_-=\Omega$, and this means that $\omega$ should be a $(k,k)$-form.

The analogues of \cite[Eq.(B.4)--(B.6)]{katmadas} become
\begin{subequations}\label{eq:I4-props}
\begin{gather}	
	\label{eq:I4'I4'} I_4'(I_4'(\rho)) = -16\, I_4(\rho)^2\, \rho\\
	\label{eq:75}  I_4'(I_4'(\rho), I_4'(\rho), \rho)= 8\, I_4(\rho) I_4'(\rho) \ ,\qquad 
	I_4'(I_4'(\rho),\rho,\rho)= -8\, I_4(\rho) \rho\\
	\label{eq:proj}
 I_4'(I_4'(\rho), \rho, \omega)=  2\, (I_4'(\rho), \omega)\, \rho + 2\,(\rho, \omega) \,I_4'(\rho) \ .
\end{gather}	
\end{subequations}
We will show these properties in appendix \ref{app:I4prop}.


\subsection{Polyform language} 
\label{sub:poly}

Using the definitions of the previous subsection, we can reformulate the flow equations of subsection \ref{sub:4d} in terms of a pure spinor on the K\"ahler base $M_6$. We therefore consider
\begin{equation}
\phi = e^{\bg + i\jg}
\end{equation}
where $\bg$, $\jg$, are the B-field and K\"ahler forms on $M_6$, as defined in subsection \ref{sub:4d}. For this case, we find the relation
\begin{equation}
 (\phi,\bar{\phi}) = - 8\,\im\,\ee^{-K}\,,
\end{equation}
where $\ee^{-K}$ is the volume defined in \eqref{eq:K-vl-def}. We define a normalised pure spinor as
\begin{equation}
\cV = \frac1{\sqrt{8\,\ee^{-K}}}\, \phi = \frac1{\sqrt{8\,\ee^{-K}}}\, e^{\bg + i\jg} \,, 
\end{equation}
for convenience in connecting with four-dimensional supergravity in the next section. One can now define a new polyform variable encompassing the K\"ahler moduli and axions through $\cV$ and the scale factor $\ee^{\chi}$ as
\begin{align}\label{eq:H-def}
\cH = &\, 2\, \ee^{\chi}\, \Im(\ee^{-i \alpha} \cV ) \,,
\end{align}
The additional phase $\ee^{i \alpha}$ is a priori arbitrary, but is fixed by our M-theory reduction as
\begin{equation}
 \ee^{i \alpha} = \gamma - \mathrm{i}\,\sqrt{1-\gamma^2}\,,
\end{equation}
with $\gamma$ as in \eqref{eq:gamma-def}. Listing separately the 0-, 2-, 4- and 6-form parts, we find
\begin{equation}
 \cH = \tfrac1{\sqrt{2}}\,\ee^{\chi} \ee^{K/2}\left( \gamma\,, \quad \gamma\, \cg\,, \quad \tfrac1{2}\gamma\, \cg \wedge \cg -\tfrac1{2}\gamma^{-1} \, \jg \wedge \jg\,, \quad f_0 \right)^{T} \,,
\end{equation}
where $f_0$ is the six-form given in \eqref{eq:F0-def}. Note that we use the combination $\cg \equiv \bg + M\,\Delta\,\jg$ defined in \eqref{eq:Lambda-def}, for brevity.

Similarly, we define two more polyforms, $\Gamma$, containing the charges, and $P$, containing the gauging in four dimensions, given by 
\begin{align}\label{eq:ch-P-k}
\Gamma \equiv &\, \tfrac1{\sqrt{2}}\,\left( p^0 + \pg + \qg + q_0\,\ee^{-K}\vl_3 \right)\,, \nonumber\\
P \equiv &\, \tfrac1{\sqrt{2}}\,\ee^{2\phi} \mg  + \tfrac1{\sqrt{2}}\,( e_0\ee^{2\phi} + 2\,E )\,\ee^{-K}\vl_3 \,, \\
k \equiv &\, \tfrac1{\sqrt{2}}\,\mg  +  \tfrac1{\sqrt{2}}\,e_0\,\ee^{-K}\vl_3 \,, \nonumber
\end{align}
where we also defined an additional polyform, $k$, for future convenience.

Using now (\ref{eq:I4*}), the flow equations \eqref{eq:0-form}, \eqref{eq:2-form}, \eqref{eq:4-form} and \eqref{eq:6-form} can be assembled into the single polyform equation
\begin{equation}\label{eq:fullflow}
 \ee^{\chi+U} \partial_r \cH  =  \frac14\, I^\prime_4(\cH,\cH,P) + \Gamma \,,
\end{equation}
while the additional scalar flow equations \eqref{eq:dil-flow} and \eqref{eq:scal-flow} take the form
\begin{align}\label{eq:phiflow}
\partial_r \ee^{-2\phi} = &\,- 2\,\ee^{-(U+\chi)}\, (\cH , k) 
\\ 
\label{eq:chiUflow}
\partial_r \left( \ee^{\chi+U} \right) = &\, - (\cH , P) \,.
\end{align}

The final conditions to be imposed are the constraints \eqref{eq:axion-con}, which can be written in a rather compact form by drawing some inspiration from their
counterparts in four-dimensional supergravity that will be described in the next section. Starting from \eqref{eq:axion-con-1}, one can verify that it is equivalent to the condition
\begin{equation}
 {\cal A} + \alpha^\prime - 2\, \ee^{-U} (\hat \cH , P ) =0\,,
\end{equation}
where $\hat \rho$ is the imaginary part of the pure spinor defined in \eqref{eq:H-def}. Here, ${\cal A}$ is defined in terms of $\cH$ as
\begin{equation}
 {\cal A} + \alpha^\prime = \tfrac12\, \ee^{-2\,\chi} \, (\cH , \partial_r\cH )\,,
\end{equation}
so that it matches with the definition of the K\"ahler connection for vector multiplet scalars in a four-dimensional supergravity theory. Similarly, \eqref{eq:axion-con-2} leads to the condition
\begin{equation}\label{eq:Aflow}
{\cal A} + \alpha^\prime = -2\, E\, \ee^{-K} \gamma^{-1}  M\,.
\end{equation}
Finally, we consider the constraint \eqref{eq:ch-con-1}, which takes the form
\begin{equation}
 (\Gamma,k)=0\,.
\end{equation}
in terms of the objects defined in (\ref{eq:ch-P-k}). Using \eqref{eq:ch-con-2} it is simple to show that $(\Gamma,P)$ is then identified with the integer $n$, appearing in that relation.

This concludes the reformulation of the relevant flow equations in terms of polyforms. We have repackaged all the flow equations as (\ref{eq:fullflow}), (\ref{eq:phiflow}), (\ref{eq:chiUflow}) (\ref{eq:Aflow}). In the next section, we will see how these equations are formally identical to the flow equations for four-dimensional black holes.

\section{Comparison with four dimensions}\label{sec:4dsugra}

In this section, we compare the flow equations obtained in the previous section to the flow equations in four-dimensional supergravity. We establish a formal equivalence, and we comment on the conceptual differences.

\subsection{Four-dimensional flows} 
\label{sub:4dflow}

In this subsection we give some details on the structure of the BPS equations for black
holes in four-dimensional gauged supergravity theories and discuss the relation to the
higher dimensional flow equations presented above. The flow equations for static,
asymptotically AdS$_4$, $1/4$-BPS black holes were derived in 
\cite{meessen-ortin, halmagyi-petrini-zaffaroni-bh} for generic models involving vector
and hyper multiplets.

We therefore consider the BPS flow equations given in \cite{halmagyi-petrini-zaffaroni-bh},
which describe solutions with metric of the type
\begin{equation}
ds_{4}^2= - \ee^{2U} dt^2 + \ee^{-2U} dr^2 + \ee^{2\chi}\, ds^2(S^2) \,,
\end{equation}
which is the four-dimensional metric one obtains upon dimensional reduction of \eqref{eq:metr-2}
along $\theta$ and $M_6$. The relevant variable for the vector multiplet scalars is the
section, $\cV$, which can be written in components in terms of scalars $X^I$ as
\begin{equation}\label{eq:sym-sec}
\cV=\begin{pmatrix} X^I\\ F_I\end{pmatrix}\,, \qquad
F_I= \frac{\partial F}{\partial X^I}\,,
\end{equation}
where $I=0,\ldots, n_v$.  $F$ is a holomorphic function of degree two, called the prepotential,which we will always consider to be cubic:
\begin{equation}\label{prep-def}
F=-\frac{1}{6}c_{ijk}\frac{X^i X^j X^k}{X^0} \,,
\end{equation}
for completely symmetric $c_{ijk}$, and now $i=1,\dots n_v$. The section $\mathcal{V}$ is subject to the constraint
\begin{equation}
  \label{eq:D-gauge}
  \Iprod{\bar{\mathcal{V}}}{\mathcal{V}} =  i  \,,
\end{equation}
where $<\,, >$ stands for the symplectic inner product. As defined here, the section $\mathcal{V}$ is
uniquely determined by the physical scalar fields, $t^i\!\equiv\!{X^i}/{X^0}$, up to a local U(1) transformation. The K\"{a}hler potential is given by
\begin{equation} 
\ee^{-K} = i \,\frac{1}{6}\,c_{ijk}(t^i-\bar t^i)(t^j-\bar t^j)(t^k-\bar t^k) \,.
\end{equation}

The real and imaginary parts of the section $\cV$ are not independent, but are related by
\begin{equation}
\Re(\cV) = 8\,I_4^\prime(\Im(\cV))\,,
\end{equation}
where we used the so called quartic invariant function $I_4(\Gamma)$, which is a quartic function of a vector, $\Gamma$, taking values in ${\mathbb R}^{2 n_v+2}$ (just like the real and imaginary parts of $\cV$); the prime denotes differentiation with respect to the argument. For symmetric cubic models described by \eqref{prep-def}, $I_4$ is a quartic polynomial: in terms of the central charge $Z(\Gamma)$ and of its K\"{a}hler covariant derivative $Z_i(\Gamma)\equiv D_{i}Z(\Gamma)$, one has
\begin{align}
I_4(\Gamma)
         =&\, - (p^0 q_0 + p^i q_i)^2 + \frac{2}{3} \,q_0\,c_{ijk} p^i p^j p^k- \frac{2}{3} \,p^0\,c^{ijk} q_i q_j q_k 
             + c_{ijk}p^jp^k\,c^{ilm}q_lq_m  
\CR
=&\, \left( Z \, \bar Z - Z_i \, \bar Z^i \right)^2 
- c_{mij} \bar Z^i \bar Z^j \, c^{mkl} Z_k  Z_l 
+  \frac{2}{3} \, \bar Z \, c^{ijk} Z^i Z^j Z^k  + \frac{2}{3} \, Z \,c_{ijk} \bar Z^i \bar Z^j \bar Z^k \,,
\label{I4-def}
\end{align}
where in the second line we omitted the argument, $\Gamma$, in all central charges for brevity.
In this case, $I_4$ satisfies various interesting properties, including \eqref{eq:I4-props}. For the more general case of homogeneous models, there is no closed expression of $I_4$ in terms of the components of the argument $\Gamma$. However, there exists an extension of the second definition in \eqref{I4-def}, in terms of special geometry invariants \cite{deWit:1992wf,D'Auria:2007ev}, which is a degree four homogeneous rational function of the central charges, but not a polynomial as in \eqref{I4-def}.

Here, we will consider generic models and use the second derivative denoted as
\begin{equation}
I^{\prime\prime}_4(\Gamma, \Gamma) = 2\, \frac{\partial^2 I_4(\Gamma)}{\partial\Gamma\,\partial\Gamma}\,.
\end{equation} 
We write two arguments $\Gamma$ to stress that for symmetric models this can be promoted to a quadratic form; for a more general model, this is only defined as written. Using the definition of $I_4$ , we find the identity
\begin{equation}
\tfrac12\, I_4(\Gamma, 2 \Im\cV, 2 \Im\cV)
    = 8\,\Iprod{\Im\!(\cV)}{\Gamma}\,\Im\cV
    +16\,\Iprod{\Re\!(\cV)}{\Gamma}\,\Re\cV
    -2\,\mathrm{J}\,\Gamma
      \,.
    \label{I4toJ}
\end{equation}
where $\mathrm{J}$ denotes the symplectic complex structure defined in terms of the vector multiplet couplings, for any $\cN=2$ supergravity. Note that this holds for both the symmetric case in \eqref{I4-def} and for the more general models, since the additional non-polynomial terms drop out when evaluated on the symplectic section itself.

With these data, we can recast the flow equations for the vector multiplet scalars and the metric scale factors, $\ee^U$, $\ee^\chi$, presented in \cite{halmagyi-petrini-zaffaroni-bh}, as
\begin{gather}
	\label{eq:4dfullflow}
 \ee^{U+\chi} \partial_r\cH = \tfrac14\,I^\prime_4(\cH,\cH,P) +\Gamma \,,
\\
\label{eq:4dUflow}
 \partial_r\ee^{U+\chi} = \Iprod{P}{\cH}\,, 
\\
\label{eq:4dAflow}
{\cal A} + \alpha^\prime + 2\, \ee^{-U}\Re\Iprod{\ee^{-\im\alpha}\cV}{P} =0 \,,
\end{gather}
where the variable $\cH$ is given by
\begin{equation}\label{eq:H-def-2}
 \cH \equiv 2\,\ee^{\chi}\,\Im(\ee^{-\im\alpha}\cV) \,.
\end{equation}
The symplectic vector, $P=(P^I, P_I)$, is related to the moment maps of the hypermultiplet sector, which describe the gauging of the theory. Here, we focus on models including a single hypermultiplet, in line with the simplifying assumption of frozen complex structure moduli in the previous sections, but the extension to add more hypermultiplets is straightforward. For the case at hand, the SU(2) triplet of moment maps $P^x$, $x=1,2,3$, can be truncated to a single vector $P^3\equiv P$, with $P^1=P^2=0$, by allowing only the dilaton to be nontrivial, while the remaining three scalars of the hyper multiplet vanish. This corresponds to our choice of trivial $B$ field along the non-compact directions and vanishing three-form modes in the ansatz of the previous section. The resulting BPS flow equation for the dilaton, $\ee^{\phi}$, reads
\begin{equation}\label{eq:4dphiflow}
\partial_r\ee^{-2\phi} =  -2\,\ee^{-U-\chi}\Iprod{\cH}{k}\,,
\end{equation}
where $k$ is the Killing vector associated to $P$, which in the case at hand is given by
\begin{equation}
 k = \ee^{-2\phi}\, \partial_\phi P \,.
\end{equation}

Finally, we must impose two global constraints arising from the spherical symmetry, one ensuring that the Killing spinor be constant over the sphere, and one coming from the fermionic sector of the hypermultiplets. These can be written as
\begin{equation}
 \Iprod{\Gamma}{P}=n \in \mathbb{Z}\,, \qquad  \Iprod{\Gamma}{k}=0\,,
\end{equation}
respectively. Note that the integer $n$ can be arbitrary, with negative values corresponding to static black holes with hyperbolic horizon, but we only consider $n=1$ in this paper for simplicity.

We can now match the equations we have obtained to those in section \ref{sub:poly}. We see that (\ref{eq:4dfullflow}), (\ref{eq:4dUflow}), (\ref{eq:4dAflow}), (\ref{eq:4dphiflow}) are formally identical to (\ref{eq:fullflow}), (\ref{eq:phiflow}) (\ref{eq:chiUflow}), (\ref{eq:Aflow}), upon viewing the various symplectic vectors of this section as the component vectors of polyforms, and identifying the Mukai pairing with the symplectic inner product as
\begin{equation}
 \Iprod{\,}{\,} \equiv (\, , \,)\,.
\end{equation}
It follows that we may identify the variables $\cH$ in \eqref{eq:H-def} and \eqref{eq:H-def-2} and take the moment map $P$ and the Killing vector $k$ of the hypermultiplet target space to be given by the expressions in \eqref{eq:ch-P-k}, while the scale factors for the metric and the dilaton on the two sides are trivially identified.


\subsection{Properties of solutions}
\label{sub:sol}

The task of finding analytic solutions to the flow equations \eqref{eq:4dfullflow}--\eqref{eq:4dAflow} and \eqref{eq:4dphiflow} is rather hard in the general case. Here, we comment on some general features of the solutions. We will first discuss the asymptotic AdS$_4$ region and the black hole AdS$_2\times$S$^2$ attractor. We will then turn to the analytical form of the solution in the case with constant dilaton \cite{katmadas,halmagyi-bh}. In the general case with flowing hypermultiplets, regular numerical example solutions exist \cite{halmagyi-petrini-zaffaroni-bh}.  

\paragraph{Asymptotic AdS$_4$} \hspace{.2cm}\\
In order to obtain the conditions at infinity, we assume constant physical scalars, while the metric functions behave as
\begin{equation}\label{eq:asymp}
 \ee^{\chi} = r + \mathcal{O}(r^0) \,, 
 \qquad
 \ee^{U} = I_4(P)^{1/4}\,r + \mathcal{O}(r^0) \,,
\end{equation}
where we used the requirement that $R_{AdS}=I_4(P)^{-1/4}$ is the radius of the asymptotic AdS$_4$. It follows that the variable $\cH={\cal A}\,r$ for some constant vector ${\cal A}$, which by \eqref{eq:4dfullflow} is
\begin{equation}
\cH = \frac12\,I_4(P)^{-3/4}\,I^{'}_{4}(P) \, r + \mathcal{O}(r^0)\,.
\end{equation}
Imposing a constant dilaton at the AdS$_4$ we also obtain its value by setting the right hand side of \eqref{eq:4dphiflow} to zero:
\begin{equation}\label{eq:dil-inf}
\Iprod{\cH}{k}=0 \quad\Rightarrow\quad \Iprod{I'_4(P)}{k}=0\,.
\end{equation}
This can be solved explicitly for the class of gaugings in \eqref{eq:ch-P-k}, as it turns out to be linear in the dilaton. The reason is that the difference $R = P - \ee^{2\phi} k=\sqrt{2}\,E \,\ee^{-K}\vl_3$ is a vector with a single component, a so-called \emph{very small} vector, satisfying the properties
\begin{equation}
 I_4(R) = I'_4(R) = 0\,, \qquad I_4(R,R,\Gamma_1,\Gamma_2) = - \Iprod{R}{\Gamma_1}\,\Iprod{R}{\Gamma_2}\,,
\end{equation}
for any $\Gamma_1$, $\Gamma_2$, by definition. The condition \eqref{eq:dil-inf} can now be solved as
\begin{equation}\label{eq:dil-val}
 3\,\Iprod{I'_4(k)}{R} + 4\,I_4(k)\,\ee^{2\phi}=0 \quad\Rightarrow\quad \ee^{2\phi}=\frac{3\,\Iprod{R}{I'_4(k)}}{4\,I_4(k)}\,,
\end{equation}
which can be evaluated for any particular model. From the M-theory point of view, this condition translates to the requirement that the internal manifold $M_6$ is K\"ahler--Einstein, so that the $S^1$ fibration over it is Sasaki--Einstein.

\paragraph{Attractor geometries} \hspace{.2cm}\\
The other interesting point is at the horizon of the extremal black hole described by the BPS flows, where all physical scalars are again constant \cite{Gutowski:2011xx,Gutowski:2012eq}. The geometry is now AdS$_2\times S^2$ and the various fields behave as
\begin{equation}
\ee^\chi = \ee^{\chi_0} + \mathcal{O}(r)\,,
\qquad
\ee^U = \ee^{U_0} r + \mathcal{O}(r^2) \,,
\end{equation}
where $\chi_0$, $U_0$ are constants, so that $\cH$ in \eqref{eq:H-def-2} is a constant vector. We then evaluate \eqref{eq:4dfullflow}--\eqref{eq:4dAflow} and \eqref{eq:4dphiflow} to obtain the following set of algebraic equations for the values of the scalar fields at the attractor point:
\begin{subequations}\label{eq:4datt}
	\begin{gather}
		\label{eq:4dattr}
	\tfrac14\,I^\prime_4(\cH,\cH,P) +\Gamma =0 \,,
	\\
	\label{eq:4dUatrr}
	\Iprod{P}{\cH}=\Iprod{k}{\cH}=0\,.
	\end{gather}	
\end{subequations}
Here $P$ is now understood to contain the constant value of the dilaton, which is to be found by \eqref{eq:4dUatrr}, once \eqref{eq:4dattr} is solved for $\cH$ in terms of $P$ and $\Gamma$.

\paragraph{Solutions with constant dilaton} \hspace{.2cm}\\
It is interesting to point out that the subset \eqref{eq:4dfullflow}--\eqref{eq:4dAflow} for a constant dilaton can be integrated in the general case \cite{halmagyi-bh}. Here, we point out that some of these solutions can be embedded in the system above, by arranging that \eqref{eq:4dphiflow} is trivially satisfied. 

In order to ensure a constant dilaton, one must set to zero the quantity $\Iprod{\cH}{k}$ in \eqref{eq:4dphiflow}, in which case the remaining equations become identical to the ones in \cite{halmagyi-bh} for a constant gauging equal to $P$ in \eqref{eq:ch-P-k}. Turning this around, we may simply consider the solution for general constant gauging and evaluate the additional condition of vanishing $\Iprod{\cH}{k}$, so that we obtain a constrained set of solutions embedded in the theory including the dilaton. The solution of \cite{halmagyi-bh} is expressed in terms of a polynomial with vector coefficients as
\begin{equation}\label{eq:hal-sol}
\ee^{U+\chi}\cH = \frac{6}{\sqrt{I_4(P)}}\,I'_4(P)\,r^3 + A_2\,r^2 + A_1\,r \,;
\end{equation}
the explicit expressions for $A_1$ and $A_2$ in terms of $P$ and of the charge $\Gamma$ can be found in \cite[Sec.~3.1]{halmagyi-bh}. A constant dilaton solution to the flow equations of the previous section is obtained by setting to zero the inner product of $k$ with each of the vectors appearing in \eqref{eq:hal-sol}. The first is trivially satisfied, since it is the boundary condition for the AdS$_4$ vacuum at infinity \eqref{eq:dil-inf}, so that it provides the constant value for the dilaton \eqref{eq:dil-val}. The remaining conditions represent two nontrivial constraints that can be interpreted as restricting the possible charge vector 
\begin{equation}\label{eq:cnst-dil}
\Iprod{A_1}{k}=\Iprod{A_2}{k}=0\,,
\end{equation}
upon using the explicit expressions in \cite{halmagyi-bh}.

Whether such solutions are realised depends on the regularity of the horizon for the charges restricted by \eqref{eq:cnst-dil}, or in other words, by the compatibility of the value \eqref{eq:dil-val} for the dilaton at infinity with the system of equations \eqref{eq:4dattr}-\eqref{eq:4dUatrr}. In order to illustrate this more concretely, we consider the class in \cite{katmadas}, for which the K\"ahler phase is constant and \eqref{eq:cnst-dil} take the simple form
\begin{equation}\label{eq:const-cond}
 I_4(\Gamma,\Gamma,\Gamma,P)=0\,, \qquad I_4(\Gamma,\Gamma,P,k)=0\,.
\end{equation}
Both of these are linear in the dilaton, so that they can be compared to \eqref{eq:cnst-dil}, resulting in
\begin{equation}\label{eq:ch-cond-restr}
 -\frac{\Iprod{I'_4(\Gamma)}{R}}{\Iprod{I'_4(\Gamma)}{k}} = \frac{3\,\Iprod{R}{I'_4(k)}}{4\,I_4(k)}\,,
 \qquad
 -\frac{I_4(\Gamma,\Gamma,R,k)}{I_4(\Gamma,\Gamma,k,k)} = \frac{3\,\Iprod{R}{I'_4(k)}}{4\,I_4(k)}\,.
\end{equation}
For any given model of the class we consider, specified by $R$ and $k$, the conditions \eqref{eq:ch-cond-restr} can be solved explicitly in terms of the components of $\Gamma$.

\subsection{Lifting to eleven dimensions} 
\label{sub:lift}

Formally, the eleven-dimensional flow equations of section \ref{sub:poly} and the four-dimensional flow equations of the previous subsection are identical. This raises the hope that one could solve them by using the same strategies used to solve the flow equations in four dimensions, which we just reviewed in the previous subsection. 

One should be careful, however, to distinguish between the $I_4$ in four-dimensional supergravity and the one we used in our M-theory approach. To stress the difference, let us call these $I_4^{\rm sugra}$ and $I_4^{\rm Hit}$ respectively. 

These two are not exactly the same. A first difference is that $I_4^{\rm Hit}$ as defined in (\ref{eq:Hita}) makes sense for any forms $\alpha_i$, whether in cohomology or not, whereas $I_4^{\rm sugra}$ is a function of the charges, which are in cohomology.\footnote{Remember, however, that the charges should also satisfy (\ref{eq:ch-con}).} One can then consider the restriction of $I_4^{\rm Hit}$ on the cohomology; this is now a space of finitely many parameters. However, even this is not exactly the same as $I_4^{\rm sugra}$. The reason can be seen by going back to the definition (\ref{eq:Hit}), (\ref{eq:Hita}); it contains terms of the type $(\alpha_1, \Gamma_{AB} \alpha_2)$. Since there is no integral over $B_6$ in this expression, each of the entries of this 12$\times$12-dimensional matrix is a function on $B_6$, not a constant. Hence $I_4$ will in general not be a number, but a function on $B_6$. Thus, even the restriction of $I_4^{\rm Hit}$ to cohomology is not the same as $I_4^{\rm sugra}$.\footnote{\label{foot:kk}A perhaps more intrinsic way of phrasing this is the following. One can divide the matrix $(\alpha_1, \Gamma_{AB} \alpha_2)$ in four 6$\times$6 blocks, just like in (\ref{eq:Jblock}), according to whether the indices $A$ and $B$ describe a vector or a one-form. For example, the block $a_{mn}=(\alpha_1, \iota_{\del_m}\iota_{\del_n} \alpha_2)$ will be a two-form; there will also be a bi-vector block $b^{mn}$, and blocks $c^m{}_n$, $d_m{}^n$, sections of $T \otimes T^*$. In terms of these blocks, $I_4=-\frac16(a_{mn}b^{mn}+c^m{}_n d_m{}^n)$. We can now expand each of the blocks in a basis on $B_6$; the two-form $a_{mn}$, for example, will be a sum over all the possible two-forms on $B_6$. There is no reason a priori that this sum should truncate to only the terms in cohomology. A similar logic applies to the other terms in the sum. Thus, even if the entries $\alpha_1,\ldots, \alpha_4$ are harmonic, evaluating $I_4^{\rm Hit}(\alpha_1,\alpha_2,\alpha_3,\alpha_4)$ involves non-harmonic forms and tensors.} We can say that in a sense $I_4^{\rm Hit}$ involves higher Kaluza--Klein modes, while $I_4^{\rm sugra}$ does not. Physically, we expect this difference to be related to the black hole being smeared or localized in the internal directions. 

These considerations make it harder than it might seem to solve the flow equations of section \ref{sub:poly}. We will not fully analyse their properties in this paper; what follows is a preliminary analysis. 

The simplest case is $M_6$ being a coset $G/H$. In this case, the M-theory reduction on $M_7$ (the U(1) fibration over $M_6$), which is a coset as well, was worked out in \cite{cassani-koerber-varela}; it turns out to be a consistent truncation, and it results in an ${\cal N}=2$ gauged supergravity. In general, such a reduction proceeds via identifying a certain finite set of forms that are closed under the exterior $d$ and the Hodge $*$. These are not always easy to find, but in the coset case a natural candidate is given by \emph{left-invariant} forms. Evaluating $I_4^{\rm Hit}$ on these forms should not involve higher Kaluza--Klein modes. (In the language of footnote \ref{foot:kk}, the two-form $a_{mn}$ will be itself an invariant two-form, and so on.) So in this case our formalism recovers the M-theory solutions that one would obtain by uplifting the four-dimensional supergravity solutions using the fact that the reduction of \cite{cassani-koerber-varela} is a consistent truncation. 

In more general situations, the situation is less clear. Recall first from section \ref{sub:bh-ansatz} that our $M_6$ is assumed to be a K\"ahler--Einstein manifold of positive curvature at infinity and remains K\"ahler along the flow. On such an $M_6$, which is not a coset, we expect that the higher Kaluza--Klein modes will indeed appear into $I_4^{\rm Hit}$, and considerably complicate the task of showing that solution exist. 

Let us first think about an attractor solution. In that case, the relevant equations are (\ref{eq:4datt}). Already solving $(P,{\cal H})=0$ looks like a challenge, since it contains a wedge product which is not an integral. Suppose however we can solve it, and let us move on (\ref{eq:4dattr}). Let us take $\Gamma$ to be in the cohomology of $M_7$. (In a reduction one also includes forms that are not in the cohomology, but the components of $\Gamma$ along those would be related to massive vectors, and would not be associated to conserved charges.) These are forms in the cohomology of $M_6$ such that (\ref{eq:ch-con}) holds. In general, even if ${\cal H}$ and $P$ are in cohomology, $I_4^{\rm Hit}{}'$ is not necessarily in cohomology, for the same reasons discussed above for $I_4^{\rm Hit}$. However, if ${\cal H}$ is the real part of a closed pure spinor, the situation simplifies a bit. If for simplicity we set the $B$ field (namely, the axions) to zero, we have ${\cal H}= {\rm Re}(e^{i \theta} e^{i \jg})$, where $\jg$ is a K\"ahler form. Now, (\ref{eq:I4'J}) (applied to the case $\rho={\cal H}$, $\omega=P$) contains $\jg \wedge \omega$ and $\jg \llcorner \omega$; if $\jg$ is K\"ahler and $\omega$ is in cohomology, both these forms will be in cohomology as well (this is the famous Lefschetz ${\rm Sl}(2,{\mathbb R})$ action on the cohomology of a K\"ahler manifold). So the left hand side of (\ref{eq:4dattr}) is in cohomology. However, $\Gamma$ is  not just closed: it is even harmonic. We can try to show that the left hand side of (\ref{eq:4dattr}) is harmonic by using (\ref{eq:I4*}), which we can do since $P$ is a sum of $(k,k)$-forms. This contains $(P,\hat {\cal H})$; if we can arrange for this to be constant on $M_6$, we have then shown that the left-hand side of (\ref{eq:4dattr}) is harmonic, and we have reduced (\ref{eq:4dattr}) to a finite-dimensional equation. Unfortunately, just like $(P,{\cal H})=0$, also $(P,\hat {\cal H})={\rm const}.$ is hard because of the absence of an integral in the pairing $(\,,)$ (recall its definition (\ref{eq:pairing})). Indeed, the presence of equalities involving wedges of forms without integrals was one of the key assumptions in the above-mentioned reduction on cosets \cite[Sec.2.2]{cassani-koerber-varela}, which is precisely the case which we previously argued to work.

Thus, already finding solutions in the attractor limit is non-trivial. For the full flow, the problems look still harder. Given that currently explicit solutions are only known in the case with constant dilaton, one would have to first impose that condition. At this point one might hope to use the general formulas in \cite{halmagyi-bh}, replacing everywhere $I_4^{\rm Hit}$ for $I_4^{\rm sugra}$. This would however requires a long series of properties (see App.~A.3 of that paper) that we have not proved to be valid in general for $I_4^{\rm Hit}$. Another option is to also assume that the K\"ahler phase is constant. To impose both this and the constant dilaton, we have to satisfy (\ref{eq:cnst-dil}); then one can use the solutions in \cite{katmadas}, which assume a smaller set of properties of $I_4$; these are (\ref{eq:I4prop}) and (\ref{eq:I4-props}), which we prove in appendix \ref{app:I4prop}. The problem is once again that (\ref{eq:cnst-dil}) are non-trivial to satisfy; in fact, these are even harder than (\ref{eq:4dattr}), where at least two of the entries was one of the two pure spinors defining the geometry of $M_6$.

In spite of all these difficulties, we think that the formal similarities between the black hole flow equations for four-dimensional supergravity and for M-theory are strong enough that they suggest the existence of black hole solutions for a general K\"ahler--Einstein $M_6$ of positive curvature. Such solutions have not been found before: the M-theory reduction on the $M_7$ obtained as $S^1$-fibrations over $M_6$ have not been worked out in general, and thus for general $M_6$ there is no known relationship with any four-dimensional effective Lagrangian. In view of our results, it would be interesting to work out such a reduction, to find more conclusive evidence for the existence of our black holes. This would presumably happen adapting to eleven dimensions the formalism in \cite{grana-louis-waldram}, probably taking into account some of the caveats in \cite{kashanipoor-minasian}. In the formalism of those papers, one needs a ``special'' basis of forms, closed under the exterior differential $d$ and the Hodge star $*$, but not necessarily harmonic. A natural candidate on $M_7$ is simply given by the pullback of the harmonic forms on $M_6$, which are not all harmonic after the pullback. 

Such a program is also interesting in view of the recent surge of results in K\"ahler--Einstein manifolds with positive curvature. Beyond the cosets mentioned earlier, it was once a bit hard to produce examples; it required some application of the continuity method \cite{tian,nadel} or in some limited setting the solution of certain ODEs \cite{koiso-sakane}. In the toric case, the existence of a K\"ahler--Einstein metric is equivalent to the barycenter of the toric polytope being the origin \cite{wang-zhu,mabuchi}.

More recently, the old Yau--Tian--Donaldson conjecture has been proven \cite{chen-donaldson-sun}: it relates the existence of a K\"ahler--Einstein metric to an algebraic-geometrical condition called K-stability. While this might condition might seem hard to implement for practical examples, it has already yielded some concrete results: for example the proof \cite{datar-szekelyhidi} of the existence of a K\"ahler--Einstein metric on certain threefolds with 2-torus action \cite{suess}, which generalize toric manifolds.


\section*{Acknowledgements}
We would like to thank J.~Stoppa, A.~Zaffaroni for interesting discussions. S.K.~and A.T.~are supported in part by INFN and by the European Research Council under the European Union's Seventh Framework Program (FP/2007-2013) -- ERC Grant Agreement n. 307286 (XD-STRING). The research of A.T.~is also supported by the MIUR-FIRB grant RBFR10QS5J ``String Theory and Fundamental Interactions''.

\appendix

\section{Some properties of the Hitchin functional} 
\label{app:I4prop}

In this appendix we are going to derive the crucial properties (\ref{eq:I4prop}) of the Hitchin functional $I_4$. 

Let us start by deriving some preliminary results. First, notice that
\begin{equation}
	(\phi, \Gamma_{AB} \phi)=0 \ .
\end{equation}
This holds because $\phi$ only has non-zero pairing with $\bar \phi$, and two gamma matrices are not enough to turn $\phi$ into $\bar \phi$. (For more details on this type of logic, see for example \cite[Sec.~2.1]{t-reform}.) Recalling now that $\rho= {\rm Re}  \phi$, $\hat \rho= {\rm Im} \phi$, we obtain 
\begin{equation}\label{eq:rgr}
	(\hat \rho , \Gamma_{AB} \hat \rho)= {\cal Q}_{AB}\ ,\qquad (\rho, \Gamma_{AB}\hat \rho)=  0
\end{equation}
Now, by definition we get $I_4'(\hat \rho,\hat \rho,\hat \rho)= 2\,{\cal Q}_{AB} \Gamma^{AB}= 6\, (\hat \rho, \rho) \rho$. Using (\ref{eq:I4'r}) we get (\ref{eq:I4'I4'}). Very similar steps also show (\ref{eq:75}). 

The identity \eqref{eq:I4*} is harder to obtain. Let us first introduce a ket--bra notation: 
\begin{equation}
	\alpha(\beta,\cdot) \equiv | \alpha \rangle \langle \beta | \ .
\end{equation}
We will make extensive use of the Fierz identity\footnote{The numerical factors in (\ref{eq:fierz}) and (\ref{eq:ABAB}) are slightly unusual; this is due to the fact that the Clifford algebra ${\Gamma_A, \Gamma_B}= {\cal I}_{AB}$ is also slightly unusual, as it misses a factor of 2.}
\begin{equation}\label{eq:fierz}
	| \alpha \rangle \langle \beta | = \frac1{64}\sum_{k=1}^{12} \frac{2^k}{k!} (\beta, \Gamma_{A_1\ldots A_k} \alpha) \Gamma^{A_k\ldots A_1}\ .
\end{equation}
We will also need the formula
\begin{equation}\label{eq:ABAB}
	\Gamma_{AB} \Psi_k \Gamma^{AB} = (3-(6-k)^2)\Psi_k\ ,
\end{equation}
where $\Psi_k$ is a bispinor of degree $k$. Here we mean an element of the tensor product space of two Clifford$(6,6)$ spinors, namely of two differential forms. This is a $64\times64$-dimensional space; one should not get confused by the fact that a single differential form can also be viewed as a bispinor for ordinary Clifford$(6)$ spinors on $M_6$. 

As a warm-up, let us apply this formalism to a pure spinor $\phi=\rho+ i\hat \rho$. If we consider $|\phi\rangle\langle \phi |$ in (\ref{eq:fierz}), all the bilinears $(\phi,\Gamma_{A_1\ldots A_k}\phi)=0$ except when $k=6$. (In some contexts this is even given as a definition of pure spinor.) So 
\begin{equation}
	|\phi\rangle\langle \phi |= \frac1{6!}(\phi, \Gamma_{A_1\ldots A_6} \phi) \Gamma^{A_6\ldots A_1}\ .
\end{equation}
 If we apply (\ref{eq:ABAB}) to this we get
\begin{equation}\label{eq:GffG}
	\Gamma_{AB} |\phi\rangle\langle \phi | \Gamma^{AB}= 3 |\phi\rangle\langle \phi |\ . 
\end{equation}
As a check, we can multiply this from the right by $\rho$; we get ${\cal Q}_{AB} \Gamma^{AB} \phi= 3i(\hat \rho,\rho)\phi$, which is essentially (\ref{eq:phirho}). 

Let us now apply the same method to $|\rho\rangle \langle \rho|$. Among the bilinears $(\rho, \Gamma_{A_1\ldots A_k} \rho)$, all those with odd $k$ vanish by chirality. The case $k=0$ vanishes because the pairing (\ref{eq:pairing}) is antisymmetric; the case $k=4$ vanishes because
\begin{equation}
	(\rho, \Gamma_{ABCD} \rho) = (\Gamma_{DCBA} \rho,\rho)= - (\rho,\Gamma_{DCBA}\rho)=-(\rho,\Gamma_{ABCD}\rho)\ .
\end{equation}
This leaves us only with the $k=2,6,10$ bilinears. The $k=10$ bilinear can actually be related to the $k=2$ case with the help of the chirality operator $\Gamma=\Gamma_1 \ldots \Gamma_{12}$. Moreover, the $k=2$ bilinear is nothing but $(\rho, \Gamma_{AB}\rho)={\cal Q}_{AB}$. All in all (\ref{eq:fierz}) gives 
\begin{equation}
	| \rho \rangle \langle \rho | = -\frac1{32} {\cal Q}_{AB} \Gamma^{AB} (1+ \Gamma)+ 
	\frac1{6!} (\rho, \Gamma_{A_1\ldots A_6} \rho) \Gamma^{A_6\ldots A_1}\ .
\end{equation}
If we now use (\ref{eq:ABAB}) on this, we get
\begin{equation}
\begin{split}
	\Gamma_{AB} |\rho\rangle \langle \rho| \Gamma^{AB} &=
	\frac{13}{32}  {\cal Q}_{AB} \Gamma^{AB} (1+ \Gamma)+\frac3{6!}(\rho, \Gamma_{A_1\ldots A_6} \rho) \Gamma^{A_6\ldots A_1}\\
	&= \frac12{\cal Q}_{AB} \Gamma^{AB} (1+ \Gamma) + 3 | \rho\rangle\langle \rho |\ .
\end{split}	
\end{equation}
Using now (\ref{eq:I4''}) and the definition of $I_4'$ it is easy to obtain (\ref{eq:I4'J}).

Finally, using (\ref{eq:GffG}) we also obtain
\begin{equation}
	\Gamma_{AB} |\rho \rangle \langle \hat \rho | \Gamma^{AB} +
	\Gamma_{AB} |\hat\rho \rangle \langle \rho | \Gamma^{AB} = 
	3\left( | \rho \rangle \langle \hat \rho | + | \hat \rho \rangle \langle \rho |\right)\ .
\end{equation}
Using this, (\ref{eq:rgr}), and the definition (\ref{eq:I4'}), one obtains (\ref{eq:proj}). 


\section{The Noether potential}\label{app:noe-pot}

Consider a generic Lagrangian ${\cal L}$ in $D$ dimensions that depends on
fields that we collectively call $\phi$ and their derivatives. Assuming general covariance,
a diffeomorphism along a vector $\xi^\mu$ induces the following transformation on the Lagrangian
\begin{equation}
\delta_\xi{\cal L}= \partial_\mu (\xi^\mu{\cal L})\,,
\end{equation}
On the other hand, one can perform a general variation of the action to obtain the equations of motion $E$,
up to a boundary term linear in the field variations $\delta\phi$, that we indicate by $\theta^\mu$
\begin{equation}
\delta{\cal L}= E\,\delta\phi + \partial_\mu \theta^\mu(\delta\phi)\,.
\end{equation}
When the generic variation is assumed to be a diffeomorphism, the two expressions must coincide:
\begin{equation}
\partial_\mu (\xi^\mu{\cal L}) = E\,\delta\phi + \partial_\mu\theta^\mu(\delta_\xi\phi)\,.
\end{equation}
It then follows that there exists a current associated with any field
configuration:
\begin{equation}\label{eq:Noe-curr}
J^\mu=\theta^\mu(\delta_\xi\phi)-\xi^\mu{\cal L} \quad\Rightarrow\quad
\partial_\mu J^\mu=-E\,\delta\phi\,,
\end{equation}
which is conserved when the configuration is a solution to the equations
of motion. This is known as the Noether current associated to the diffeomorphism generated by $\xi^\mu$. 
As shown in \cite{wald-coh}, any conserved current locally constructed from fields can be
written as the divergence of an antisymmetric tensor, using the equations of motion. It follows
that one can locally define the so called Noether potential through
\begin{equation}\label{eq:Noe-pot}
J^\mu= \partial_\nu Q^{\mu\nu} \,,
\end{equation}
which also depends linearly on $\xi^\mu$. 

The existence of these objects allows for a definition of a charge associated with backgrounds for which $\xi^\mu$
is a symmetry. This passes through the definition of a generator of symmetries on the space of all solutions viewed
as a manifold (i.e. the phase space), the so called symplectic current
\begin{equation}
\Omega^\mu(\delta\phi, \delta_\xi\phi) = \delta\theta^\mu(\delta_\xi\phi) - \delta_\xi\theta^\mu(\delta\phi)\,,
\end{equation}
and is identified with the variation of the corresponding Hamiltonian associated with the symmetries. In the case
that $\xi^\mu$ is a symmetry of the solution at hand, $\delta_\xi\phi$ and consequently $\Omega$ vanish identically,
reflecting the existence of an irrelevant, or pure gauge, direction in the solution space. This current can
be computed by variation of \eqref{eq:Noe-curr}, as
\begin{gather}
   \Omega^\mu(\delta\phi, \delta_\xi\phi)  = \delta J^\mu -
\Pi^\mu_\xi\,,\label{eq:gener-from-J}\\
   \Pi^\mu_\xi \equiv \delta_\xi\theta^\mu(\delta\phi) - \delta (\xi^\mu{\cal L})\,.
\end{gather}
In this paper we consider only diffeomorphisms $\xi^\mu$ along rotational Killing vectors, which correspond to angular momentum. By the requirement that the cycles used in the various integrals are invariant under the rotational Killing vectors, it turns out that the integral of $\Pi^\mu_\xi$ over any spatial section vanishes, so we will disregard its presence in the following.

The definition of the conserved charge can be given by computing the integral of \eqref{eq:gener-from-J} over the total spatial manifold $\Sigma$ as
\begin{equation}\label{J-int}
\int_{\Sigma} \Omega = \delta \int_{S_1} Q - \delta \int_{S_2} Q = 0
\,,
\end{equation}
where we used the Gauss theorem and $S_{1,2}$ are $D-2$-dimensional spacial hypersurfaces. In the last equality, we 
imposed that $\xi^\mu$ is a symmetry, so that $\Omega^\mu$ vanishes. The conserved charge can then be defined through
\begin{equation}\label{eq:Noe-charge}
\mathcal{Q}= \int_{S} Q \,,
\end{equation}
which is independent of the hypersurface. 

We now briefly specialise these ideas to the case of a Lagrangian describing a gauge three-form interacting with gravity
through terms at most quadratic in derivatives, assuming that the Lagrangian does not contain any bare gauge fields.
After a diffeomorphism and a general variation of the Lagrangian, one finds
\begin{align}
\theta^\mu(\delta\phi) = \,
2\, ({\cal L}_G^{\mu\nu\rho\sigma}\nabla_{\rho}\delta g_{\sigma\nu} -\nabla_{\rho}{\cal L}_G^{\rho\nu\mu\sigma}\delta g_{\sigma\nu})
+ \, 2\,{\cal L}_F^{\mu\nu\rho\sigma}\delta A_{\nu\rho\sigma} \,,
\end{align}
where we defined the derivatives of the Lagrangian with respect to the four-form field strength and the Riemann tensor as
\begin{equation}
{\cal L}_F^{\mu\nu\rho\sigma}= \frac{\partial{\cal L}}{\partial F_{\mu\nu\rho\sigma}} \,,\quad
{\cal L}_G^{\mu\nu\rho\sigma} = \frac{\partial{\cal L}}{\partial R_{\mu\nu\rho\sigma}} \,.
\label{notation}
\end{equation}
Using these results, Noether potential reads
\begin{eqnarray}\label{Noe-diffs}
Q^{\mu\nu}   &=&   2\,{\cal L}_G^{\mu\nu\rho\sigma}\nabla_{\rho}\xi_\sigma-4\,\nabla_{\rho}{\cal L}_G^{\mu\nu\rho\sigma}\xi_\sigma 
                +\, 2\,(\xi^\lambda A_{\lambda\rho\sigma})\,{\cal L}_F^{\mu\nu\rho\sigma} \,.\nonumber
\end{eqnarray}
This expression can be used in \eqref{eq:Noe-charge} to obtain a conserved charge associated to a rotational isometry $\xi^\mu$.
In this paper, we apply this formalism to the slightly more involved case of the bosonic sector of eleven dimensional supergravity,
which does contain bare gauge fields. Nevertheless, the procedure above can be followed in exactly the same way, to obtain \eqref{eq:J-def}
in the case of a rotational Killing vector, defining an angular momentum charge.

\bibliography{at} \bibliographystyle{JHEP}
 
\end{document}